\documentclass{article}[11pt]
\usepackage{a4wide}
\usepackage{bm}
\usepackage{bbm}
\usepackage{epsfig,graphics}
\usepackage{amsmath,amssymb,amsfonts}
\usepackage{color}
\usepackage{epstopdf}
\usepackage{slashed}
\usepackage[%dvipdfm,
colorlinks=true,
linkcolor=black,
breaklinks=true,
urlcolor=blue,
citecolor=green]{hyperref}

\usepackage[font=footnotesize, labelfont=bf]{caption,subcaption}
\usepackage{booktabs}
\usepackage{dcolumn}
\usepackage{makecell}
\usepackage{multirow}
\newcolumntype{d}{D{.}{.}{5}}

\usepackage[square,comma,sort&compress,numbers]{natbib}

\newcommand{\be}{\begin{equation}}
\newcommand{\ee}{\end{equation}}
\newcommand{\ba}{\begin{eqnarray}}
\newcommand{\ea}{\end{eqnarray}}

\newcommand{\Lag}{\mathcal{L}}

\newcommand{\eps}{\epsilon}

\newcommand{\mM}{\mathcal{M}}

\begin{document}

\title{Decay behaviors of the $P_c$ hadronic molecules}
\date{\today}

\author{Yong-Hui Lin$^{1,2,}$\footnote{Email address:
      \texttt{linyonghui@itp.ac.cn} }~ ,
      Chao-Wei Shen$^{1,2,}$\footnote{Email address:
      \texttt{shencw@itp.ac.cn} }~ ,
      Feng-Kun Guo$^{1,2,}$\footnote{Email address:
      \texttt{fkguo@itp.ac.cn} }~ and
      Bing-Song Zou$^{1,2,}$\footnote{Email address:
      \texttt{zoubs@itp.ac.cn} }
        \\[2mm]
      {\it\small$^1$CAS Key Laboratory of Theoretical Physics, Institute
      of Theoretical Physics,}\\
      {\it\small  Chinese Academy of Sciences, Beijing 100190,China}\\
      {\it\small$^2$ School of Physical Sciences, University of Chinese Academy
of Sciences, Beijing 100049, China} \\
}

\maketitle

\begin{abstract}

The $P_c(4380)$ and $P_c(4450)$ states observed recently by LHCb experiment were proposed to be either $\bar{D} \Sigma_c^*$ or $\bar{D}^* \Sigma_c$ S-wave bound states of spin parity $J^P={\frac32}^-$. We analyze the decay behaviors of such two types of hadronic molecules within the effective Lagrangian framework. With branching ratios of ten possible decay channels calculated, it is found that the two types of hadronic molecules have distinguishable decay patterns. While the $\bar{D} \Sigma_c^*$ molecule decays dominantly to $\bar{D}^* \Lambda_c$ channel with a branching ratio by 2 orders of magnitude larger than to $\bar{D}\Lambda_c$, the $\bar{D}^* \Sigma_c$ molecule decays to these two channels with a difference of less than a factor of 2. Our results show that the total decay width of $P_c(4380)$ as the spin-parity-${\frac32}^-$ $\bar{D} \Sigma_c^*$ molecule is about a factor of 2 larger than the corresponding value for the $\bar{D}^* \Sigma_c$ molecule. It suggests that the assignment of $\bar{D} \Sigma_c^*$ molecule for $P_c(4380)$ is more favorable than the $\bar{D}^* \Sigma_c$ molecule. In addition, $P_c(4450)$ seems to be a $\bar{D}^* \Sigma_c$ molecule with $J^P={\frac52}^+$ in our scheme. Based on these partial decay widths of $P_c(4380)$, we estimate the cross sections for the reactions $\gamma p \to J/\psi p $ and $ \pi p\to J/\psi p $ through the s-channel $P_c(4380)$ state. The forthcoming $\gamma p$ experiment at JLAB and $\pi p$ experiment at JPARC should be able to pin down the nature of these $P_c$ states.

\end{abstract}

\medskip

\newpage

\section{Introduction}

  In recent years, a large number of new hadrons were discovered experimentally
following the developments in the high-energy experiments and the accumulation
of the precise data in the low-energy exclusive
measurements~\cite{Olive:2016xmw}. Some of these hadrons were suggested to
have internal structure more complex than the simple $q \bar{q}$ configuration
for mesons or $qqq$ configuration for baryons in the traditional picture of the
constituent quark models, and are good candidates of exotic hadrons. Study of
exotic hadrons has become a central topic of hadron spectroscopy in the past
decade (for a recent review, see Ref.~\cite{Chen:2016qju}).
Especially,
the observation of two hidden charm pentaquark-like structures, $P^+_c(4380)$
and $P^+_c(4450)$, by the LHCb Collaboration~\cite{Aaij:2015tga} attracts much
attention. The existence of relatively narrow hidden-charm pentaquarks has been
suggested and their masses have been predicted in
Refs.~\cite{Wu:2010jy,Yang:2011wz,Yuan:2012wz,Xiao:2013yca}. The $P_c$
structures are observed in the $J/\psi p$ invariant mass distribution in the
process of $\Lambda^0_b \to J/\psi p K^-$ decay. Assuming they are resonances,
the reported mass and width of the $P^+_c(4380)$ are $(4380\pm8\pm29)
\ \mathrm{MeV}$ and $(205\pm 18\pm 86)\ \mathrm{MeV}$, respectively, while the
$P^+_c(4450)$ has a mass of $(4449.8 \pm 1.7\pm 2.5) \ \mathrm{MeV}$ and a
width of $(39\pm 5\pm 19)\ \mathrm{MeV}$. The spin-parities of these two $P_c$
states are not well determined yet. According to experimental analyses, the
most favorable set of the spin and parity for the lower and the higher peaks is
$J^P=(3/2^-, 5/2^+)$.

Many models for the structure and production of
$P_c$ states have been proposed, such as the baryon--meson
molecules~\cite{Meissner:2015mza,Chen:2015moa,He:2015cea,Roca:2015dva,Chen:2015loa, Xiao:2015fia,Lu:2016nnt,Shen:2016tzq}, compact pentaquark
states~\cite{Maiani:2015vwa,Li:2015gta,Mironov:2015ica,
Anisovich:2015cia,Ghosh:2015ksa,Wang:2015epa,Ali:2016dkf}  and
baryocharmonia~\cite{Kubarovsky:2015aaa}, while the possibility of
rescattering-induced kinematical effects has
also been discussed~\cite{Guo:2015umn,Liu:2015fea,Guo:2016bkl,Bayar:2016ftu}.
Some of these models are used to predict other possible pentaquark-like hadrons,
and others are built to explore the internal structure of the $P_c$ states.
Among them, the one which we are interested in is that the $P_c(4380)$ and
$P_c(4450)$ are interpreted as the hadronic molecular states composed of either
$\bar{D} \Sigma_c^*(2520)$ or $\bar{D}^* \Sigma_c(2455)$ since their masses are
quite close to these two thresholds. Based on this hadronic molecular picture,
we try to analyze the decay behaviors of $P_c$ states by calculating the partial
widths of the $P_c$ states into some possible final states in the framework of
effective Lagrangian approach. It will help us to distinguish different
interpretations about the structure of $P_c$ states and can be examined by
future experiments.

This work is organized as follows. After introduction, we first present the details of the theoretical formalism in Sec.~\ref{sec:formalisms}. The predicted decay properties and some discussion are presented in Sec.~\ref{sec:results}. Finally, a brief summary will be given and an appendix is presented in the end.

\section{Formalism} \label{sec:formalisms}

In this section, we present the fundamental formalism for the investigation about the decay properties of the $P_c$ states in the $\bar{D} \Sigma_c^*$ and  $\bar{D}^* \Sigma_c$ molecular pictures. The spin-parity of $P_c(4380)$ state is set to be ${\frac 3 2}^-$ in the whole work.

\begin{figure}[htbp]
\begin{center}
\includegraphics[width=0.5\textwidth]{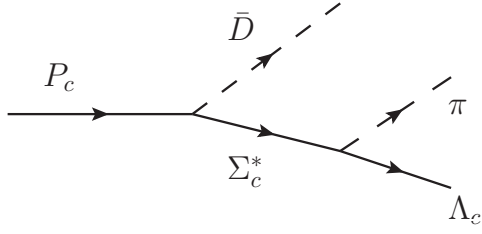}
\caption{The three-body decay of the $P_c$ state as the $\bar{D} \Sigma_c^*$ molecule.
\label{Fig: threebody}}
\end{center}
\end{figure}

Since both the $\Sigma_c^{*+}$ and $\Sigma_c^{*++}$ are unstable with a width of about $15\ \mathrm{MeV}$, and decays dominantly into the $\pi\Lambda_c$, a natural decay mode for the $\bar D \Sigma_c^*$ molecule would be the three-body $\bar D\pi\Lambda_c$, as shown in Fig.~\ref{Fig: threebody}, where final state interaction has been neglected. The decay widths of $D^*$ and $\Sigma_c$ are less than $2\ \mathrm{MeV}$. Therefore, the three- and four-body decays through the decays of the
$D^*$ and $\Sigma_c$ in the $\bar D^*\Sigma_c$ hadronic molecule can be
neglected.
%---------
%---------

\begin{figure}[tb]
\begin{center}
\includegraphics[width=0.31\textwidth]{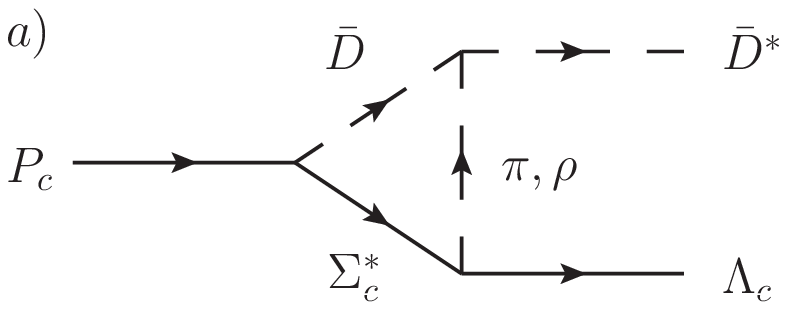}
\includegraphics[width=0.31\textwidth]{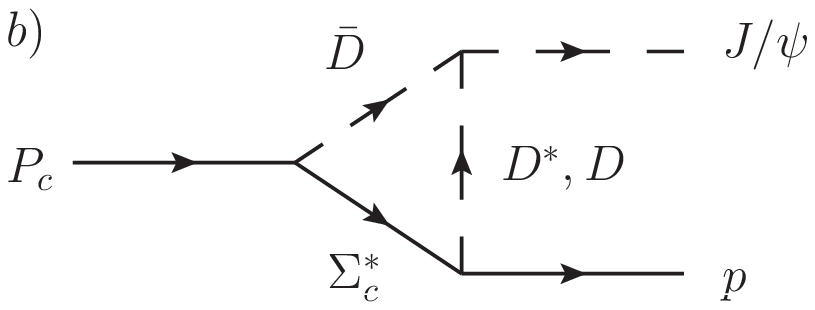}
\includegraphics[width=0.31\textwidth]{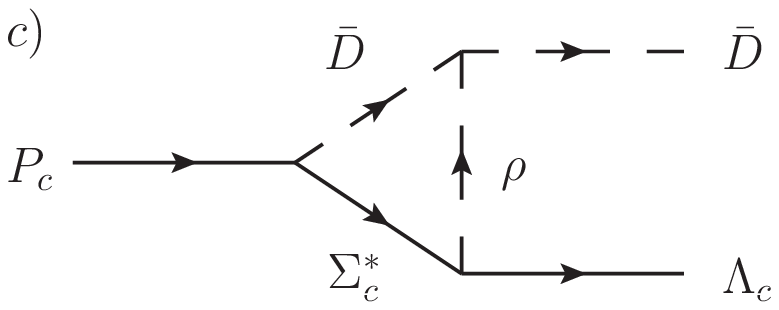}\\ \ \\
\includegraphics[width=0.31\textwidth]{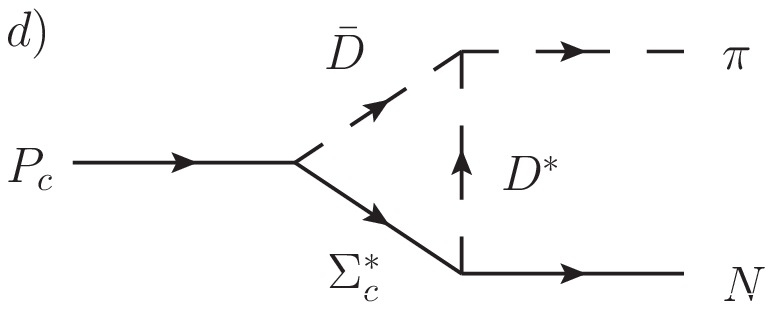}
\includegraphics[width=0.31\textwidth]{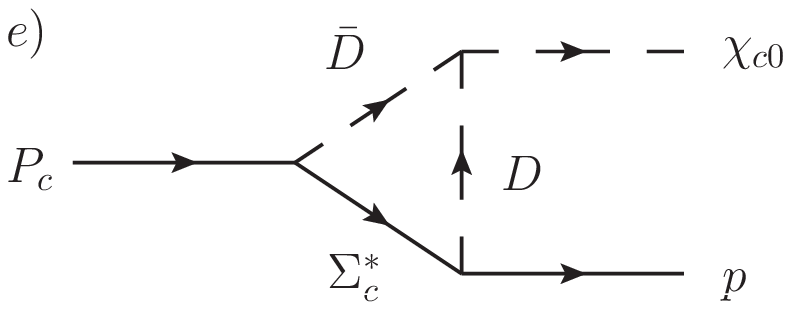}
\includegraphics[width=0.31\textwidth]{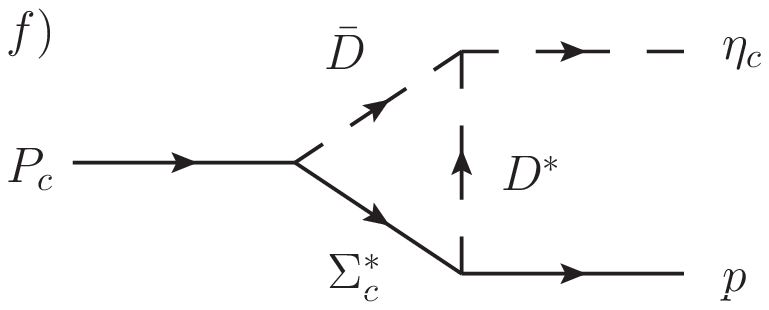}\\ \ \\
\includegraphics[width=0.31\textwidth]{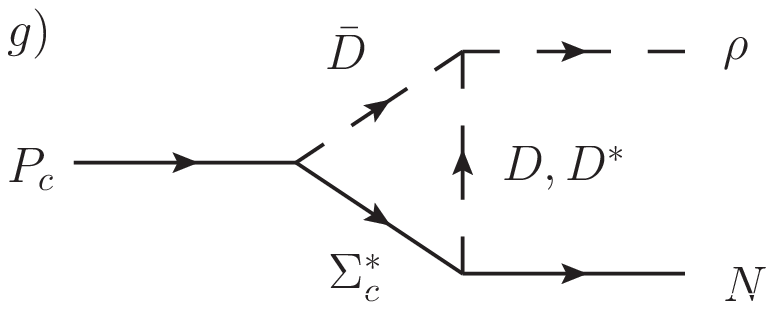}
\includegraphics[width=0.31\textwidth]{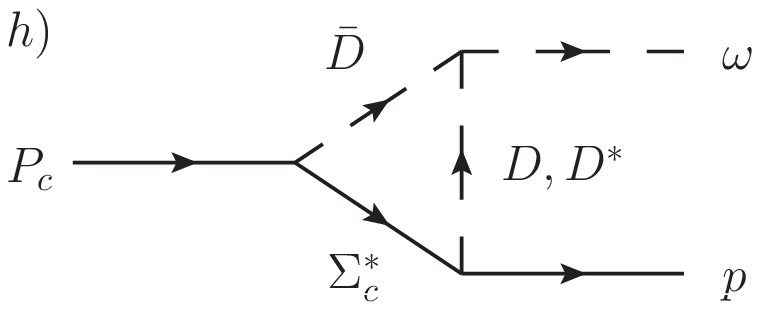}
\includegraphics[width=0.31\textwidth]{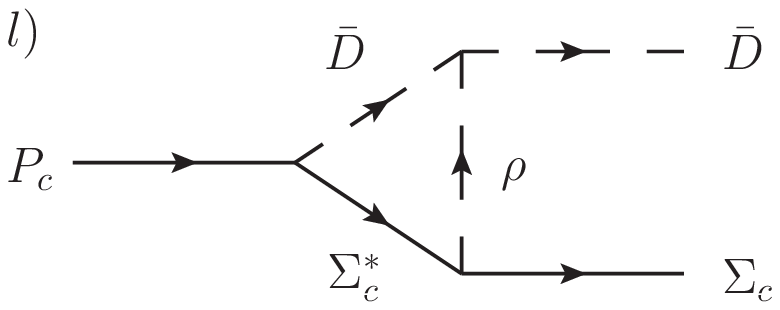}
\caption{The decays of the $P_c$ state as a $\bar{D} \Sigma_c^*$ molecule.
a) $\bar D^*\Lambda_c$ channel with $\pi$ exchange dominant and $\rho$ exchange secondary.
b) $J/\psi p$ channel with $D^*$ exchange dominant and $D$ exchange secondary.
c) $\bar D\Lambda_c$ channel with $\rho$ exchange.
d) $\pi N$ channel with $D^*$ exchange.
e) $\chi_{c0}p$ channel with $D$ exchange.
f) $\eta_c p$ channel with $D^*$ exchange.
g) $\rho N$ channel with $D$ exchange dominant and $D^*$ exchange secondary.
h) $\omega p$ channel with $D$ exchange dominant and $D^*$ exchange secondary.
l) $\bar D\Sigma_c$ channel with $\rho$ exchange.
\label{Fig: mechanism}}
\end{center}
\end{figure}

\begin{figure}[htbp]
\begin{center}
\includegraphics[width=0.31\textwidth]{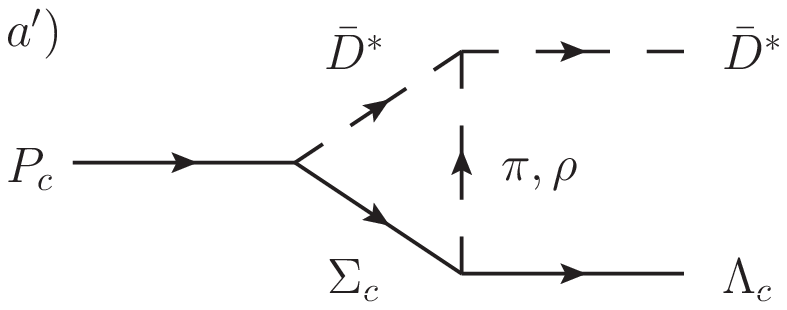}
\includegraphics[width=0.31\textwidth]{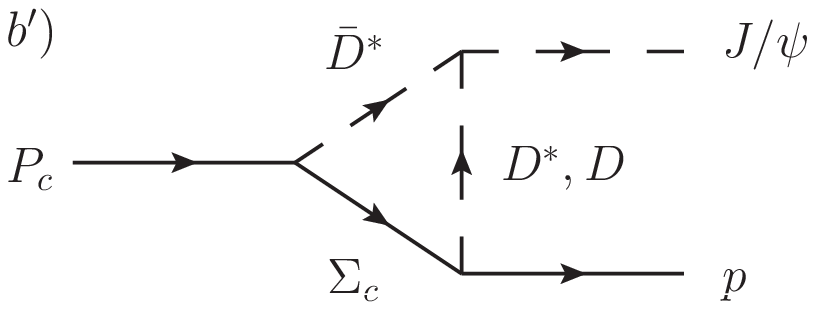}
\includegraphics[width=0.31\textwidth]{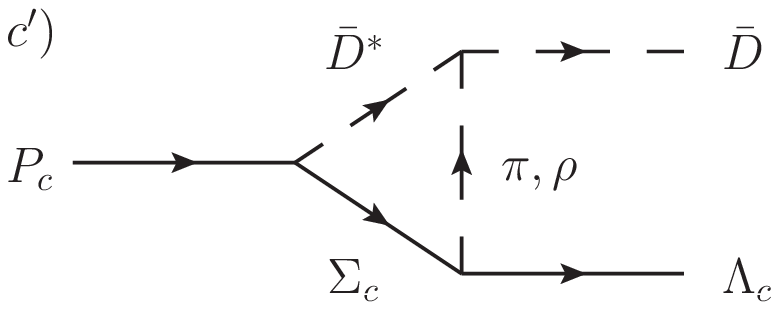}\\ \ \\
\includegraphics[width=0.31\textwidth]{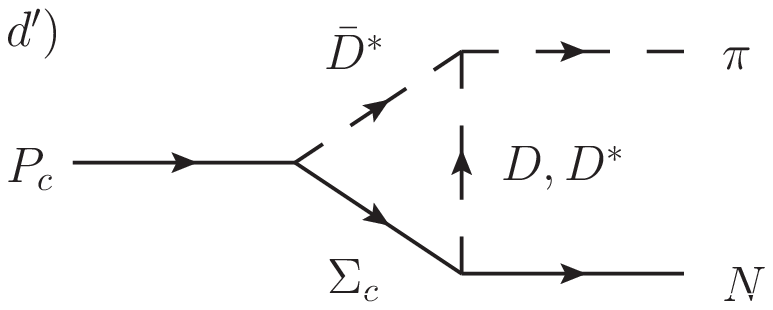}
\includegraphics[width=0.31\textwidth]{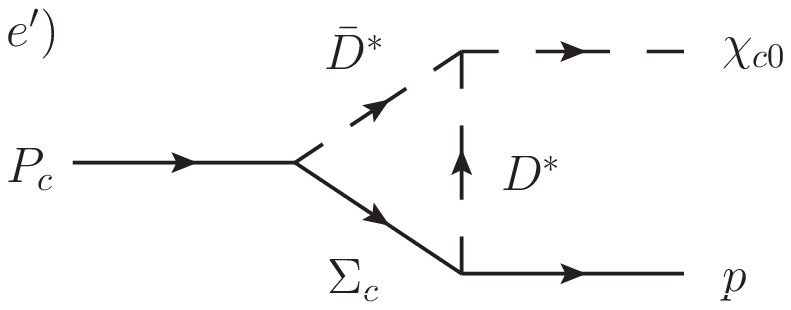}
\includegraphics[width=0.31\textwidth]{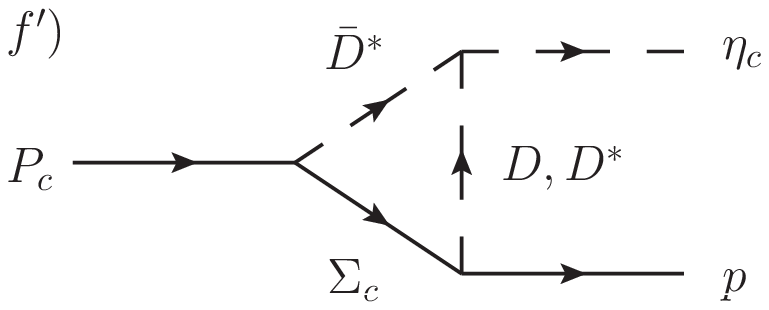}\\ \ \\
\includegraphics[width=0.31\textwidth]{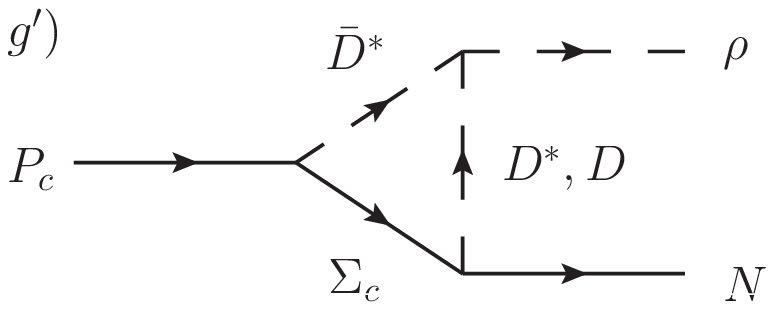}
\includegraphics[width=0.31\textwidth]{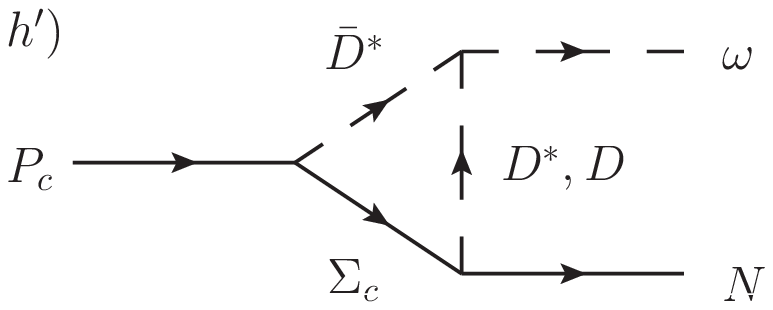}
\includegraphics[width=0.31\textwidth]{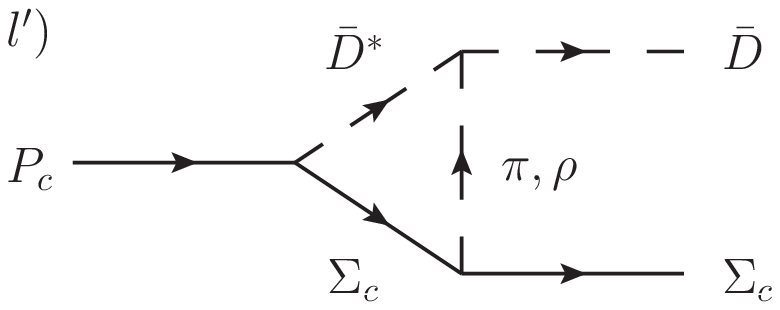}
\caption{The decays of the $P_c$ state as the $\bar{D}^* \Sigma_c$ molecule.
$a^{\prime}$) $\bar D^*\Lambda_c$ channel with $\pi$ exchange dominant and $\rho$ exchange secondary.
$b^{\prime}$) $J/\psi p$ channel with $D^*$ exchange dominant and $D$ exchange secondary.
$c^{\prime}$) $\bar D\Lambda_c$ channel with $\pi$ exchange dominant and $\rho$ exchange secondary.
$d^{\prime}$) $\pi N$ channel with $D$ exchange dominant and $D^*$ exchange secondary.
$e^{\prime}$) $\chi_{c0}p$ channel with $D^*$ exchange.
$f^{\prime}$) $\eta_c p$ channel with $D$ exchange dominant and $D^*$ exchange secondary.
$g^{\prime}$) $\rho N$ channel with $D^*$ exchange dominant and $D$ exchange secondary.
$h^{\prime}$) $\omega p$ channel with $D^*$ exchange dominant and $D$ exchange secondary.
$l^{\prime}$) $\bar D\Sigma_c$ channel with $\pi$ exchange dominant and $\rho$ exchange secondary.
\label{Fig: mechanism1}}
\end{center}
\end{figure}

Because of the small widths of the $\Sigma_c$ and $\Sigma_c^*$, these three-body decay modes are not the main contribution to the total width of the hadronic
molecules under consideration. The $P_c$ states can also decay into a meson and
a baryon. The two-body decay modes which will be considered in this paper
are listed in Table~\ref{table:final}.
%---------
\begin{table}[htpb]
\centering
\caption{\label{table:final}All possible final states for the $P_c(4380)$ and $P_c(4450)$ decay with $J^P={\frac32}^-$.}
% \scalebox{1.2}{
\begin{tabular}{l|c}
\Xhline{1pt}
\thead{Initial state} & \thead{Final states} \\
\Xhline{0.8pt}
$P_c(4380)(\bar{D} \Sigma_c^*)$ 	 & $\bar D^*\Lambda_c$, $J/\psi p$, $\bar D\Lambda_c$, $\pi N$, $\chi_{c0}p$, $\eta_c p$, $\rho N$, $\omega p$, $\bar D\Sigma_c$ 	\\
$P_c(4380)(\bar{D}^* \Sigma_c)$ 	 & $\bar D^*\Lambda_c$, $J/\psi p$, $\bar D\Lambda_c$, $\pi N$, $\chi_{c0}p$, $\eta_c p$, $\rho N$, $\omega p$, $\bar D\Sigma_c$	\\
\Xhline{0.8pt}
$P_c(4450)(\bar{D}^* \Sigma_c)$ 	 &  $\bar D^*\Lambda_c$, $J/\psi p$, $\bar D\Lambda_c$, $\pi N$, $\chi_{c0}p$, $\eta_c p$, $\rho N$, $\omega p$, $\bar D\Sigma_c$, $\bar{D} \Sigma_c^*$      \\
\Xhline{1pt}
\end{tabular}
% }
\end{table}
%---------
Note that the threshold of $\bar{D} \Sigma_c^*$ system is about $4386\
\mathrm{MeV}$, $6\ \mathrm{MeV}$ higher than the central value of the mass of
$P_c(4380)$. Thus, the $\bar{D} \Sigma_c^*$ channel appears only in the
$P_c(4450)$ decay as shown in the table. One sees that some of the decays happen
at relatively long distances, i.e., the involved momentum exchange is small,
such as the ones into a pair of anti-charm meson and charm baryon. In principal,
the partial widths of such decays could be calculated in the framework of a
nonrelativistic effective field theory. However, due to the lack of knowledge of
the interaction between the anti-charm mesons and charm baryons, we have to rely
on models such as exchanging the $\rho$ meson in addition to the lightest pion.
The decays into a pair of light meson and light baryon take place at short
distances, and again an estimate of such decays can only be done in models. The
triangle diagrams for the meson-exchange model of the two-body decays of the
$\bar D\Sigma_c^*$ and $\bar D^*\Sigma_c$ hadronic molecules are shown in
Fig.~\ref{Fig: mechanism} and Fig.~\ref{Fig: mechanism1}, respectively.
%---------
%---------

%---------
%---------

It should be mentioned that the sets of spin and parity for $(\bar{D},
\Sigma_c^*)$ and $(\bar{D}^*, \Sigma_c)$ are $(0^-, {\frac32}^+)$ and $(1^-,
{\frac12}^+)$, respectively. Thus the $P_c$ states of spin-parity ${\frac32}^-$
may be considered as $S$-wave bound states of $\bar{D} \Sigma_c^*$ or
$\bar{D}^* \Sigma_c$. Subject to the Lorentz covariant orbital-spin
scheme~\cite{Zou:2002yy}, the $S$-wave couplings for the $P_c$ with
$J^P={\frac32}^-$ with the meson--baryon pairs of interest are given by
\begin{eqnarray}
\Lag_{\bar{D} \Sigma_c^* P_c} &=& g_{\bar{D} \Sigma_c^* P_c}^{} \bar{\Sigma}_c^{* \mu} P_{c \mu} \bar{D} + H.c. , \notag \\
\Lag_{\bar{D}^* \Sigma_c P_c} &=& g_{\bar{D}^* \Sigma_c P_c}^{} \bar{\Sigma}_c P_{c \mu} \bar{D}^{* \mu} + H.c. ,
\label{eq:vertex0}
\end{eqnarray}
where two $S$-wave coupling constants $g_{\bar{D} \Sigma_c^* P_c}$ and $g_{\bar{D}^* \Sigma_c P_c}$ can be estimated by using~\cite{Weinberg:1965zz,Baru:2003qq}
\be
{g^2} = \frac{4\pi}{4 M m_2}  \frac{(m_1+m_2)^{5/2}} {(m_1 m_2)^{1/2}}
\sqrt{32\epsilon} , \label{eq:coupling}
\ee
here $M$, $m_1$ and $m_2$ denote the masses of the $P_c$ state, $\bar{D}({\bar{D}}^*)$ and $\Sigma_c({\Sigma_c}^*)$, respectively, and $\eps$ is the binding energy. The factor $\frac{1}{4 M m_2}$ in Eq.~\eqref{eq:coupling} is introduced for the normalization of two fermion fields, $P_c$ and $\Sigma_c({\Sigma_c}^*)$. Assuming the physical state in question to be a pure $S$-wave hadronic molecule, the relative uncertainty of the above approximation for the coupling constant is $\sqrt{2\mu\epsilon}\, r$ where $\mu=m_1 m_2/(m_1+m_2)$ is the reduced mass of the bound particles, and $r$ is the range of forces which may be estimated by the inverse of the mass of the particle that can be exchanged. Thus, for the $\bar D \Sigma_c^*$ and $\bar D^*\Sigma_c$ systems, $r$ may be estimated as $1/m_\rho$ and $1/m_\pi$, respectively.

Besides the vertex described as Eq.~(\ref{eq:vertex0}) introduced in the previous discussion, the following effective Lagrangian~\cite{Oh:2000qr, Erkol:2006sa, Guo:2010ak,Garzon:2015zva,Shen:2016tzq,Lu:2016nnt} need to be presented for evaluating the decay amplitudes of the Feynman diagrams shown in Fig.~\ref{Fig: threebody},~\ref{Fig: mechanism},~\ref{Fig: mechanism1},
\begin{eqnarray}
\Lag_{V P_1 P_2} &=& i g_{V P_1 P_2}^{} \left( V_{\mu} \partial^{\mu} P_1 P_2 - V_{\mu} \partial^{\mu} P_2 P_1\right),\\
\Lag_{V_1 V_2 P} &=& -g_{V_1 V_2 P}^{} \ \varepsilon^{\mu \nu \alpha \beta} \left(\partial_{\mu} V_{1 \nu}  \partial_{\alpha} V_{2 \beta}\right) P,\\
\Lag_{V_1 V_2 V_3} &=& -i g_{V_1 V_2 V_3}^{} \Bigl\{ V_1^{\mu}\left(\partial_{\mu} V_2^{\nu} V_{3 \nu} - V_2^{\nu} \partial_{\mu} V_{3 \nu}\right) + \left(\partial_{\mu} V_{1 \nu} V_2^{\nu} - V_{1 \nu} \partial_{\mu} V_2^{\nu} \right) V_3^{\mu} \nonumber \\ && \mbox{} \qquad\quad
 + V_2^{\mu}\left(V_1^{\nu} \partial_{\mu} V_{3 \nu} - \partial_{\mu} V_{1 \nu} V_3^{\nu}\right) \Bigr\},\\
\Lag_{P B_1 B_2} &=& -i g_{P B_1 B_2}^{} \bar{B}_1 \gamma_5 B_2 P + H.c.,\\
\Lag_{V B_1 B_2} &=& g_{P B_1 B_2}^{} \bar{B}_1 \gamma_{\mu} V^{\mu} B_2 + H.c.,\\
\Lag_{P B B^*} &=& g_{P B B^*}^{} \bar{B}^{* \mu} \partial_{\mu} P B + H.c., \label{eq:PBB*}\\
\Lag_{V B B^*} &=& -i g_{V B B^*}^{} \bar{B}^{* \mu} \gamma^{\nu} \gamma_5 [\partial_{\mu} V_{\nu} - \partial_{\nu} V_{\mu}] B + H.c., \label{eq:VBB*}\\
\Lag_{D D \chi_{c0}} &=& i g_{D D \chi_{c0}}^{} D D + H.c.,\\
\Lag_{D^* D^* \chi_{c0}} &=& i g_{D^* D^* \chi_{c0}}^{} D^{* \mu} D^{* \dag}_{\mu} + H.c.,
\end{eqnarray}
where $V P_1 P_2$ denotes $D^* D \pi$, $D^* D \eta_c$, $\rho D D$, $J/\psi D D$ or $\omega D D$, $V_1 V_2 P$ denotes $\rho D^* D$, $\omega D^* D$, $J/\psi D^*
D$, $D^* D^* \pi$ or $D^* D^* \eta_c$, $V_1 V_2 V_3$ denotes $D^* D^* \rho$, $D^* D^* \omega$ or $D^* D^* J/\psi$, $P B_1 B_2$ denotes $D N \Sigma_c$, $\pi
\Sigma_c \Sigma_c$ or $\pi \Sigma_c \Lambda_c$, $V B_1 B_2$ means $\rho \Sigma_c \Lambda_c$, $\rho \Sigma_c \Sigma_c$ or $D^* N \Sigma_c$, $P B B^*$ means $\pi
\Lambda_c \Sigma_c^*$ or $D N \Sigma_c^*$, and finally $V B B^*$ denotes $\rho \Sigma_c \Sigma_c^*$, $\rho \Lambda_c \Sigma_c^*$ or $D^* N \Sigma_c^*$. Another
essential part in studying the decay properties of a hadronic molecule is to estimate the coupling constants appearing in related vertices. The values of
coupling constants $g_{D^* D \pi}$, $g_{\pi \Sigma_c \Lambda_c}$ and $g_{\pi \Lambda_c \Sigma_c^*}$ are deduced from the precise experimental data of the
decay widths of $D^*$, $\Sigma_c$ and $\Sigma_c^*$. In the heavy quark limit, the $S$-wave heavy mesons $D$ and $D^*$ are in the same spin multiplet. As a result, $g_{D^* D^* \pi}$ and $g_{D^* D \pi}$ are related to each other in that limit up to a normalization factor. We will take $g_{D^* D^* \pi} = {\bar{M}_D} g_{D^* D \pi}/2$ with $\bar{M}_D$ the average mass of $D$ and $D^*$ mesons following Ref.~\cite{Oh:2000qr}. In addition, in the same limit there exist relations: $g_{D^* D \eta_c}^{} = g_2 \sqrt{m_{\eta_c}^{}} m_{D}^{}$, $g_{D^* D^* \eta_c}^{} = g_2 \sqrt{m_{\eta_c}^{}}$, $g_{D D \chi_{c0}}^{} = -\sqrt{3} g_1 \sqrt{m_{\chi_{c0}}} m_{D}^{}$, and $g_{D^* D^* \chi_{c0}}^{} = -\frac1{\sqrt{3}} g_1 \sqrt{m_{\chi_{c0}}^{}} m_{D^*}^{}$. However, because all of the ground state $S$-wave and $P$-wave charmonia are below open-charm thresholds, neither $g_1$ nor $g_2$ can be measured directly. For the numerical estimate of the partial widths, we will take the model values $g_1 = -4 \ \mathrm{GeV}^{-1/2}$ and $g_2 =  2.36 \ \mathrm{GeV}^{-3/2}$~\cite{Colangelo:2003sa}.\footnote{Note that these values are half of those in Ref.~\cite{Colangelo:2003sa} due to the difference in conventions.} The following coupling constants are taken from
Ref.~\cite{Lin:1999ad}: $g_{D^* D^* J/\psi} = g_{J/\psi D D}= 7.64$, %(Note that this value can also be obtain from $g_2$ in Ref.~\cite{Colangelo:2003sa}, and the difference between these two values is negligible.)
$g_{D^* D^* \rho} =g_{\rho D D} = 2.52$, $g_{D^* D^* \omega} = g_{\omega D D} =
-2.84$. The other coupling constants used in our work are listed in
Table.~\ref{table:constants}. One should notice that most of the these values
can only regarded as a rough estimate, which should suffice for an
order-of-magnitude estimate of the decay rates under consideration.
%---------
\begin{table}[tbh]
\centering
\caption{\label{table:constants}The coupling constants used in this paper from Refs.~\cite{Shen:2016tzq, Lu:2016nnt, Ronchen:2012eg, Aliev:2011kn}.}
% \scalebox{1.0}{
\begin{tabular}{c|*{8}{c}}
\Xhline{1pt}
\thead{Coupling \\ constants} 	 & $g_{D^* D \pi}$ & \thead{$g_{D^* D^* \pi}$ \\ $(\mathrm{GeV}^{-1})$} & $g_{\pi \Sigma_c \Lambda_c}$ & \thead{$g_{\pi \Lambda_c \Sigma_c^*}$ \\ $(\mathrm{GeV}^{-1})$} & \thead{$g_{\rho D^* D}$ \\ $(\mathrm{GeV}^{-1})$} & \thead{$g_{\omega D^* D}$ \\ $(\mathrm{GeV}^{-1})$} & \thead{$g_{J/\psi D^* D}$ \\ $(\mathrm{GeV}^{-1})$} & $g_{D N \Sigma_c}$   	\\
\Xhline{0.8pt}
\thead{Value} 	 & 8.4 & 8.65 & 19.3 & 7.46 & 2.82 & -3.18 & 8.64 & 2.69	\\
\Xhline{1pt}
\thead{Coupling \\ constants} 	 & $g_{\pi \Sigma_c \Sigma_c}$ & $g_{D^* N \Sigma_c}$  & $g_{\rho \Sigma_c \Lambda_c}$ & $g_{\rho \Sigma_c \Sigma_c}$ & \thead{$g_{D N \Sigma_c^*}$ \\ $(\mathrm{GeV}^{-1})$} & \thead{$g_{D^* N \Sigma_c^*}$ \\ $(\mathrm{GeV}^{-1})$} & \thead{$g_{\rho \Lambda_c \Sigma_c^*}$ \\ $(\mathrm{GeV}^{-1})$} & \thead{$g_{\rho \Sigma_c \Sigma_c^*}$ \\ $(\mathrm{GeV}^{-1})$}   	\\
\Xhline{0.8pt}
\thead{Value}   & 10.76 & 3.0 & -1.04 & 13.8 & 6.5 & 2.92 & 10 & 5.77  	\\
\Xhline{1pt}
\end{tabular}
% }
\end{table}
%---------

In the rest frame of the initial state, the two-body decay width can be written as
\be
{\rm d}\Gamma = \frac{F_I}{32 \pi^2} \overline{|{\cal M}|^2}
\frac{|\mathbf{p_1}|}{M^2} {\rm d}\Omega,
\ee
where ${\rm d}\Omega = {\rm d}\phi_1 {\rm d}(\cos{\theta_1})$ is the solid angle of particle 1, $M$ is the mass of the $P_c$, the factor $F_I$ from the isospin symmetry is a constant for a certain channel, and the polarization-averaged squared amplitude $\overline{|{\cal M}|^2}$ means $\frac14 \sum_\text{spin} |{\cal M}|^2$. The amplitude expressions for all of the processes shown in Figs.~\ref{Fig: mechanism} and~\ref{Fig: mechanism1} are collected in the Appendix.

Among all the triangle diagrams, some of the amplitudes, corresponding to the exchange of a pseudoscalar meson for the $D$-wave decay modes~\cite{Albaladejo:2015dsa, Shen:2016tzq}, are ultraviolet (UV) finite while the others diverge. Nevertheless, even the UV finite loops receive short-distance contributions if we integrate over the whole momentum space. We will employ the following UV regulator which suppress short-distance contributions and thus can render all the amplitudes UV finite~\cite{Dong:2008gb,Dong:2009yp,Shen:2016tzq,Lu:2016nnt}
\be
\tilde\Phi(p^2_E /\Lambda_0^2) \equiv {\rm{exp}}(-p^2_E /\Lambda_0^2),
\label{eq:regualtor}
\ee
where $p_E$ is the Euclidean Jacobi momentum, and the cutoff $\Lambda_0$ denotes a hard momentum scale which suppresses the contribution of the two constituents at short distances $\sim 1/\Lambda_0$. The value of $\Lambda_0$ should be much larger than the typical momentum in the bound state, given by $\sqrt{2\mu\epsilon}$. It should also not be too large since we have neglected all other degrees of freedom, except for the two constituents, which would play
a role at short distances. We thus vary the value of $\Lambda_0$ from $0.5\ \mathrm{GeV}$ to $1.2\ \mathrm{GeV}$ for an estimate of the two-body partial widths. In addition, an off-shell form factor for the exchanged meson with mass $m$, momentum $q$ chosen as Eq.~(\ref{eq:ff}) needs to be introduced, and we take the
form used in, e.g., Ref.~\cite{He:2015yva}.
\be
f(q^2) = \frac{\Lambda_1^4}{(m^2 - q^2)^2 + \Lambda_1^4},
\label{eq:ff}
\ee
The parameter $\Lambda_1$ for the off-shell form factor varies for different system, and we will vary it in the range of  $1.5\, \sim \, 2.4 \ \mathrm{GeV}$~\cite{He:2014nxa}.

\begin{figure}[htbp]
\begin{center}
\includegraphics[width=0.5\textwidth]{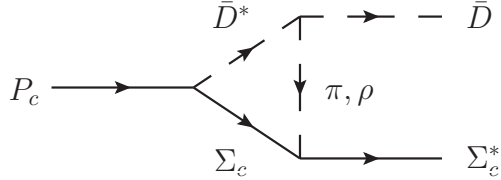}
\caption{The $\bar{D} \Sigma_c^*$ decay channel of the $P_c(4450)$ state as a
$\bar{D}^* \Sigma_c(2455)$ molecule.
\label{Fig:4450}}
\end{center}
\end{figure}

Taking all into account, we can easily get the partial widths of the $P_c(4380)$
decaying into all possible final states in both $\bar D\Sigma_c^*$ and $\bar
D^*\Sigma_c$ molecular pictures. Partial decay widths of the $P_c(4450)$ state
as a $\bar{D}^* \Sigma_c$ molecule with $J^P=3/2^-$ can also be obtained. Except
for the decay modes shown in Fig.~\ref{Fig: mechanism1}, the $\bar{D}
\Sigma_c^*$ mode is also allowed kinematically, and the diagram corresponding to
this channel is shown in Fig.~\ref{Fig:4450}.
%---------
%---------

%\newpage

\section{Decay widths of the $P_c$ states} \label{sec:results}

\begin{figure}[tb]
	\centering
        \includegraphics[width=0.49\textwidth]{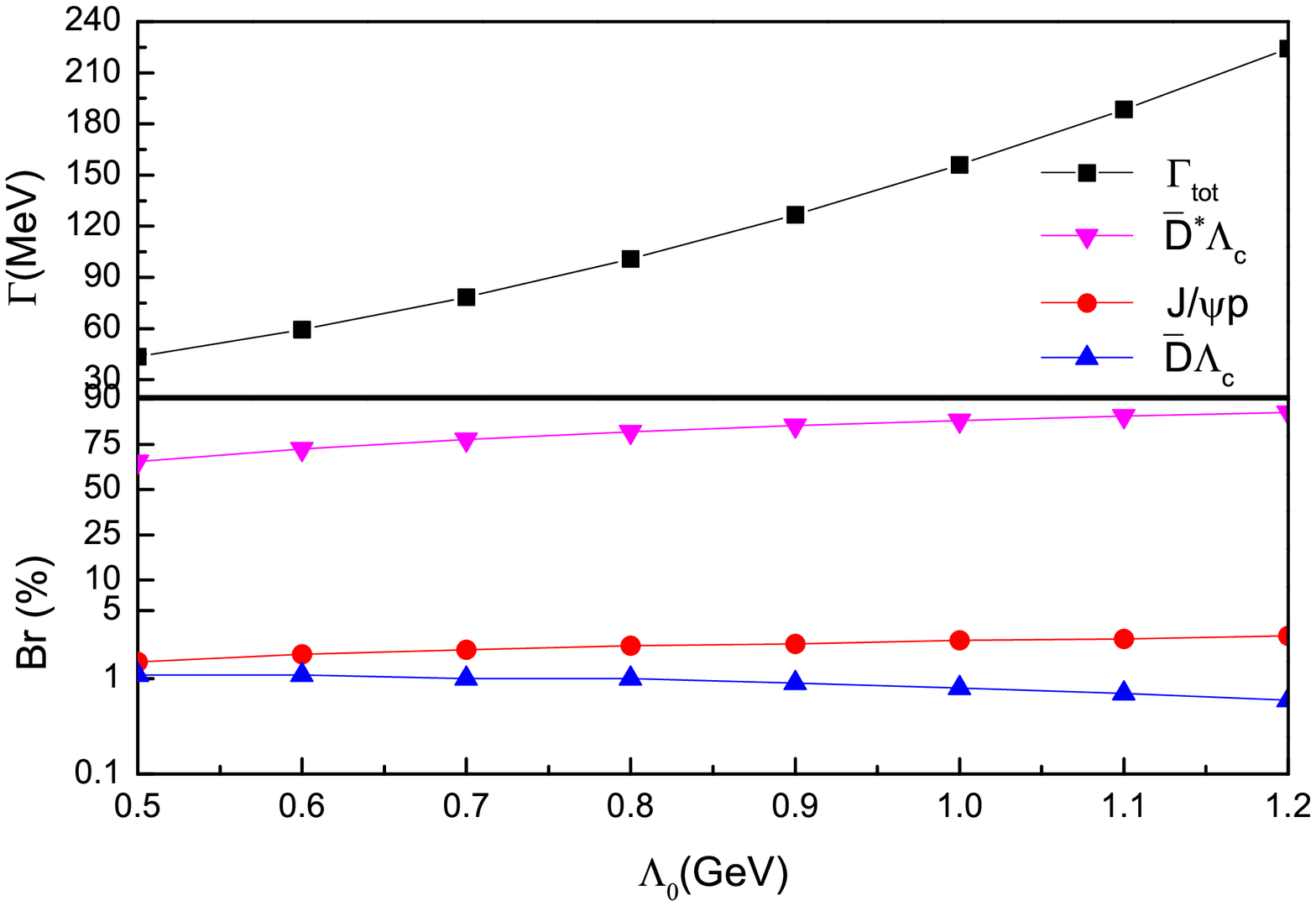}\hfill
        \includegraphics[width=0.49\textwidth]{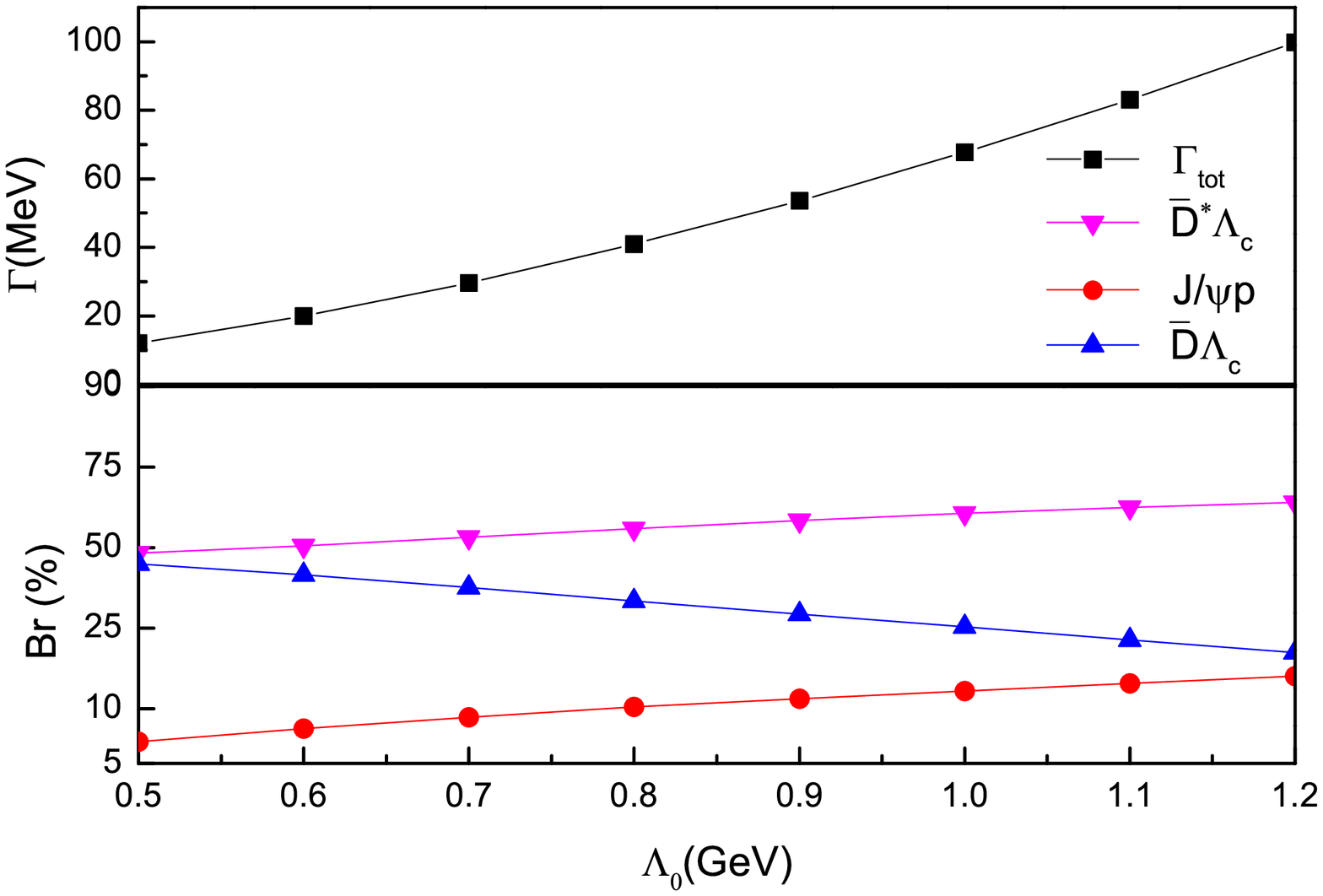}
\caption{\label{figure:comparison1}
Dependence of the $P_c(4380)$ total width and branching fractions of $\bar
D^*\Lambda_c$, $\bar D\Lambda_c$ and $J/\psi p$ on the cutoff $\Lambda_0$ in
different scenarios for the $P_c(4380)$: (a) $S$-wave $\bar D\Sigma_c^*$
molecule with $J^P={\frac32}^-$; (b) $S$-wave $\bar D^*\Sigma_c$ molecule with
$J^P={\frac32}^-$. Here $\Lambda_1$ is fixed at 2.0~GeV.
}
\end{figure}

\begin{figure}[tb]
	\centering
        \includegraphics[width=0.49\textwidth]{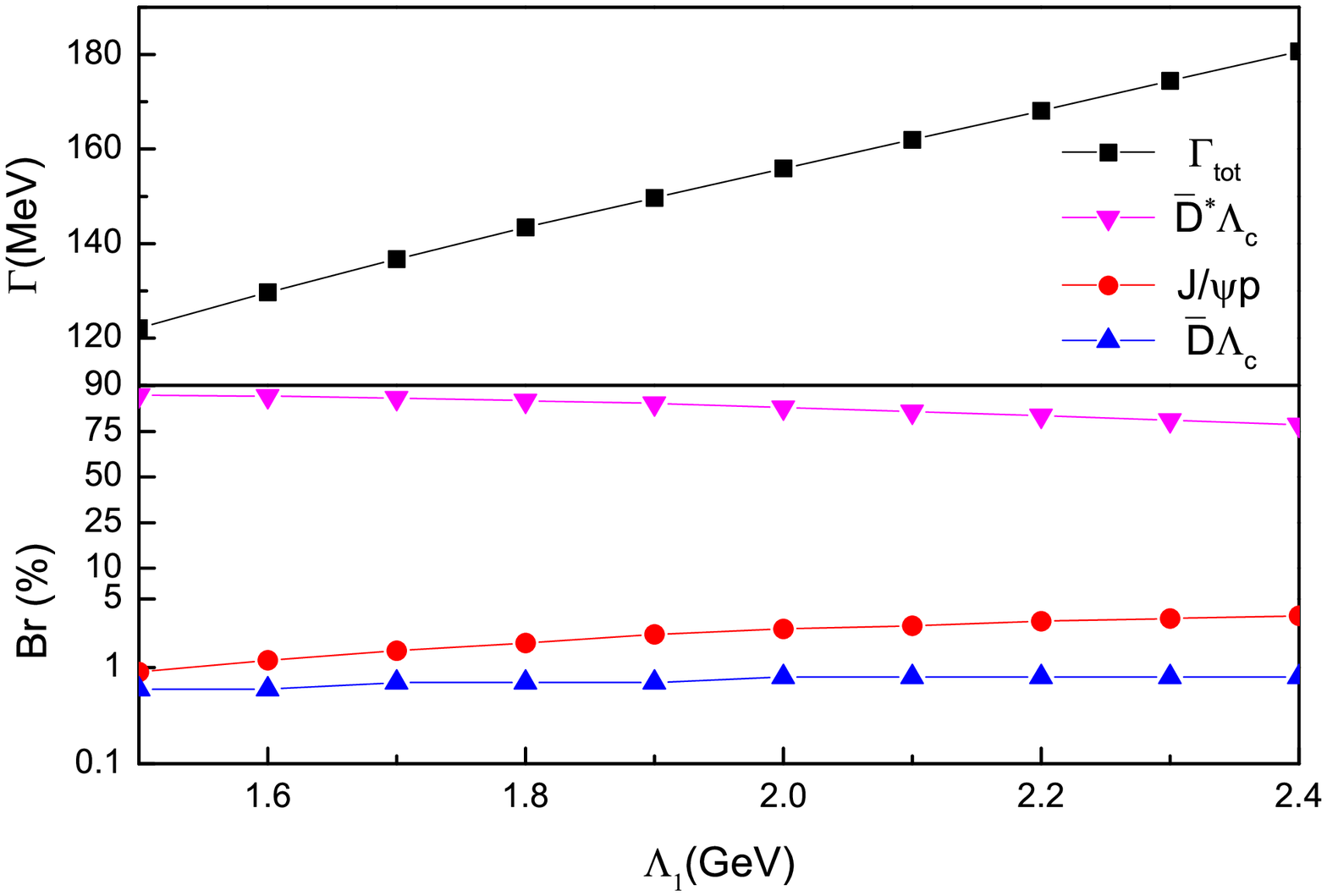}\hfill
        \includegraphics[width=0.49\textwidth]{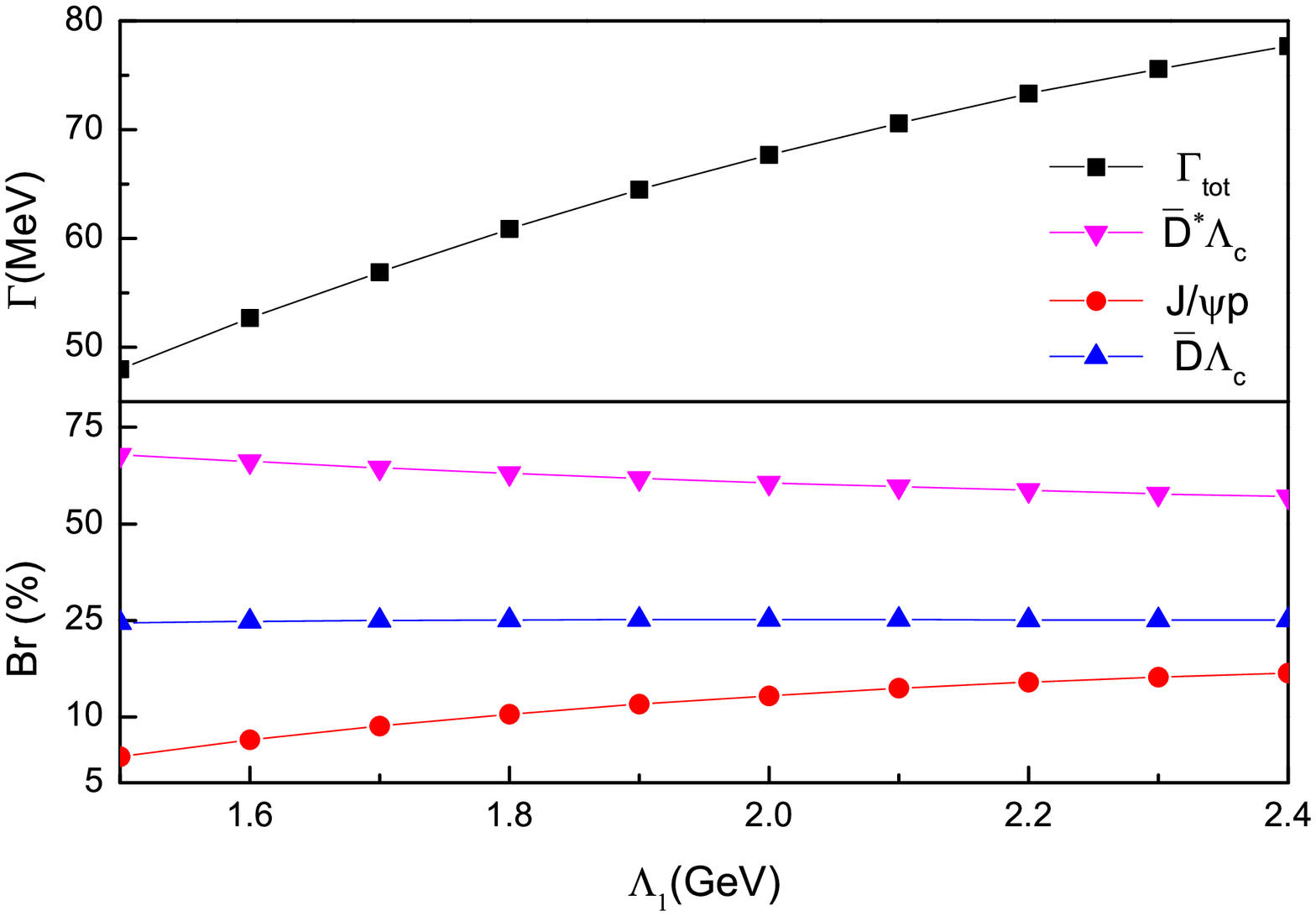}
	\caption{\label{figure:comparison} Dependence of the $P_c(4380)$ total
width and branching fractions of $\bar
D^*\Lambda_c$, $\bar D\Lambda_c$ and $J/\psi p$ on the cutoff $\Lambda_1$ in
different scenarios for the $P_c(4380)$: (a) $S$-wave $\bar D\Sigma_c^*$
molecule with
$J^P={\frac32}^-$; (b) $S$-wave $\bar D^*\Sigma_c$ molecule with
$J^P={\frac32}^-$. Here
$\Lambda_0$ is fixed at 1.0~GeV.
	}
\end{figure}

Using the values of coupling constants listed in the
Table~\ref{table:constants}, the partial decay widths of the $P_c$ states are
calculated in both $\bar{D} \Sigma_c^*$ and  $\bar{D}^* \Sigma_c$ molecular
pictures. Results obtained with typical cutoff values $\Lambda_0=1.0\
\mathrm{GeV}$ and $\Lambda_1=2.0\ \mathrm{GeV}$  are
displayed in Table~\ref{table:total}. However, these values should not be
regarded as the final results of our calculation. Our model bears a large
uncertainty due to the values of some of the coupling constants and the
choice of cutoffs $\Lambda_0$ and $\Lambda_1$.
%-----------
\begin{table}[tbh]
\centering
\caption{\label{table:total}Partial widths of $P_c(4380)$ as $\bar D \Sigma_c^*$
molecule and $\bar D^*\Sigma_c$ molecule respectively, and $P_c(4450)$ as $\bar
D^*\Sigma_c$ molecule, to different possible final states with $\Lambda_0=1.0\
\mathrm{GeV}$, $\Lambda_1=2.0\ \mathrm{GeV}$. All of the decay widths are
in the unit of $\mathrm{MeV}$, and
the short bars denote that the $\bar D\Sigma^*_c$ channel is closed in the
$P_c(4380)$ molecule decay or the corresponding contribution is negligible.}
%`` 0.00 '' denotes this value is much smaller than 0.01.}
% \scalebox{1.2}{
\begin{tabular}{l|*{4}{c}}%@{\extracolsep{1em}}*{3}{d}}
\Xhline{1pt}
\multirow{3}*{Mode} & \multicolumn{4}{c}{Widths ($\mathrm{MeV}$)} \\
\Xcline{2-5}{0.4pt}
& \multicolumn{2}{c}{$P_c(4380)$} & \multicolumn{2}{c}{$P_c(4450)$} \\
\Xcline{2-3}{0.4pt}\Xcline{4-5}{0.4pt}
& \multicolumn{1}{c}{$\bar D \Sigma_c^*$(${\frac32}^-$)} & \multicolumn{1}{c}{$\bar D^*\Sigma_c$(${\frac32}^-$)} & \multicolumn{1}{c}{$\bar D^*\Sigma_c$(${\frac32}^-$)} & \multicolumn{1}{c}{$\bar D^*\Sigma_c$(${\frac52}^+$)} \\
\Xhline{0.8pt}
$\bar D^*\Lambda_c$ 	 & 131.3  	 & 41.6   	 & 80.5 	& 22.6\\
$J/\psi p$ 		 	     & 3.8    	 & 8.4  	 & 8.3      & 2.0\\
$\bar D\Lambda_c$  	     & 1.2  	 & 17.0  	 & 41.4     & 18.8\\
$\pi N$ 			 	 & 0.06  	 & 0.05      & 0.05     & 0.1\\
$\chi_{c0}p$ 		 	 & 0.9    	 & 0.002  	 & 0.01     &
0.001 \\
$\eta_c p$ 		 	     & 0.2    	 & 0.08   	 & 0.1 	    & 0.04\\
$\rho N$ 			  	 & 1.4   	 & 0.08    	 & 0.07     & 0.1\\
$\omega p$ 		 	     & 5.3  	 & 0.3   	 & 0.3      & 0.2\\
$\bar D\Sigma_c$ 	  	 & 0.01 	 & 0.1    	 & 1.2      &
0.8\\
$\bar D\Sigma^*_c$ 	  	 & -	     & -    	 & 7.7      & 1.4\\
$\bar D\Lambda_c \pi$ 	 & 11.6   	 & -    	 & -        & -\\
\Xhline{0.8pt}
Total 				 	 & 144.3  	 & 67.7   	 & 139.7    & 46.2\\
\Xhline{1pt}
\end{tabular}
% }
\end{table}
%-----------
% In
% the $P_c(4380)$ case, the numerical results show that the total decay width
% corresponding to the $\bar{D} \Sigma_c^*$ molecule is $144.3\ \mathrm{MeV}$,
% and the value is $67.7\ \mathrm{MeV}$ for the $\bar D^*\Sigma_c$ molecule.
% Thus, the $J^P = {\frac32}^-$ $\bar{D} \Sigma_c^*$ molecular scenario seems
% more favorable than the $J^P = {\frac32}^-$ $\bar{D}^* \Sigma_c$ molecular
% scenario comparing with the experimental value.
The dependence of the $P_c(4380)$ total width on the cutoff $\Lambda_0$ in
different scenarios, together with the branching fractions of the three most
relevant channels $\bar D^*\Lambda_c$, $\bar D\Lambda_c$ and $J/\psi p$, is
shown in
Fig.~\ref{figure:comparison1}, and the dependence on
$\Lambda_1$ is shown in Fig.~\ref{figure:comparison}.
%
% Furthermore, in order to find the differences of the decay behaviors between the
% $\bar{D} \Sigma_c^*$ and $\bar{D}^* \Sigma_c$ schemes for $P_c(4380)$, we
% analyze the dependence of the partial decay widths on both $\Lambda_0$ and
% $\Lambda_1$ parameters in the form factors.
The ranges of the cutoff values are
chosen as $\Lambda_0 \in [0.5,\,1.2]\ \mathrm{GeV}$ and $
\Lambda_1 \in [1.5,\,2.4] \ \mathrm{GeV}$. It should be
mentioned that among the two-body decay modes of the $\bar{D} \Sigma_c^*$
molecule, the $\bar D^*\Lambda_c$, $J/\psi p$ and $\bar D\Lambda_c$ channels
contribute most of widths. Therefore, we only focus on these
channels for the cutoff dependence for simplicity.
%---------
%---------

The numerical results show that the $P_c(4380)$ state in both $\bar{D}
\Sigma_c^*$ and $\bar{D}^* \Sigma_c$ molecular pictures has the $\bar
D^*\Lambda_c$ as its largest decay channel. However, for the $\bar D\Lambda_c$
channel, the partial width in the $\bar{D} \Sigma_c^*$ picture for
$P_c(4380)$ is much smaller than that in the $\bar{D}^* \Sigma_c$ picture.
In addition, the relative ratio between branching fractions for the $\bar
D^*\Lambda_c$ and $J/\psi p$ channels is very different in these two kinds of
molecular scenarios. In the $J^P =
{\frac32}^-$ $\bar{D} \Sigma_c^*$ molecular picture, $\text{Br}(\bar
D^*\Lambda_c): \text{Br}(J/\psi p) \simeq 40 : 1$ when the cutoffs are fixed as
$\Lambda_0= 1.0\ \mathrm{GeV}$ and $\Lambda_1= 2\ \mathrm{GeV}$, while it is
about $ 5 : 1$ in the $J^P = {\frac32}^-$ $\bar{D}^* \Sigma_c$ picture with
these cutoffs. In particular, as one can see from Figs.~\ref{figure:comparison1}
and \ref{figure:comparison}, this conspicuous difference holds for the whole
ranges of cutoff  values that we use for $\Lambda_0$ and $\Lambda_1$. Hence,
such an interesting feature should be rather model-independent, and should be
extremely helpful for revealing the internal structure of the $P_c(4380)$ in
future experiments. Furthermore, the decay width is more sensitive to the cut
off $\Lambda_0$ in the regulators than $\Lambda_1$ in the off-shell form
factor. This is determined by their specific forms.
%---------
%---------

It is clear that the total width of the $P_c(4380)$ in the  $J^P = {\frac32}^-$
$\bar{D} \Sigma_c^*$ picture is larger than that in the $\bar D^*\Sigma_c$
picture. Thus, the former picture seems to be more consistent with the large
measured width of around 200~MeV for the $P_c(4380)$ although the latter cannot
be completely excluded given the large uncertainties of both experimental
measurements and our theoretical estimates. In any case, the dominant decay
mode of the $P_c(4380)$ is the $\bar D^*\Lambda_c$ which can proceed through
one-pion exchange. Two previous calculations~\cite{Xiao:2013yca,Ortega:2016syt}
have underestimated the partial decay width of this channel.
Ref.~\cite{Xiao:2013yca} only considered vector-meson exchanges and overlooked
the pion exchange contribution, while Ref.~\cite{Ortega:2016syt} calculated the
meson-baryon interaction from the $qq$ interaction in constituent quark model by
using the resonating group method, which sometimes fails to reproduce hadronic
observables. In fact, the pion exchange has also been found to be important for the
$\bar{D}^*\Sigma_c-\bar{D}\Sigma_c^*$ coupled channel effects~\cite{Shimizu:2016rrd,Yamaguchi:2016ote}.

The total decay width of the $P_c(4450)$ state described as
a $J^P={\frac32}^-$ $\bar{D}^* \Sigma_c$ molecule is $140\
\mathrm{MeV}$ with $\Lambda_0=1.0\ \mathrm{GeV}$ and $\Lambda_1=2.0\
\mathrm{GeV}$. This value is almost three times larger than the experimental
one reported by the LHCb. One may reproduce the experimental value by taking
the $\Lambda_0$ and $\Lambda_1$ values different from the ones used for the
$P_c(4380)$ as an $S$-wave $\bar{D}\Sigma_c^*$ molecule.
% Besides, since the two
% $P_c$ states were observed in the three body final state in $\Lambda_b$ decays,
% the rescattering-induced kinematical effects with possible triangle singularity
% cannot be fully excluded~\cite{Guo:2015umn,Liu:2015fea}.
Another possibility is that the $P_c(4450)$ has quantum numbers $J^P =
{\frac52}^+$, hence could be a P-wave $\bar{D}^* \Sigma_c$ molecule. We will
discuss this possibility in the following.

\begin{figure}[!t]
   \centering
        \includegraphics[width=0.49\textwidth]{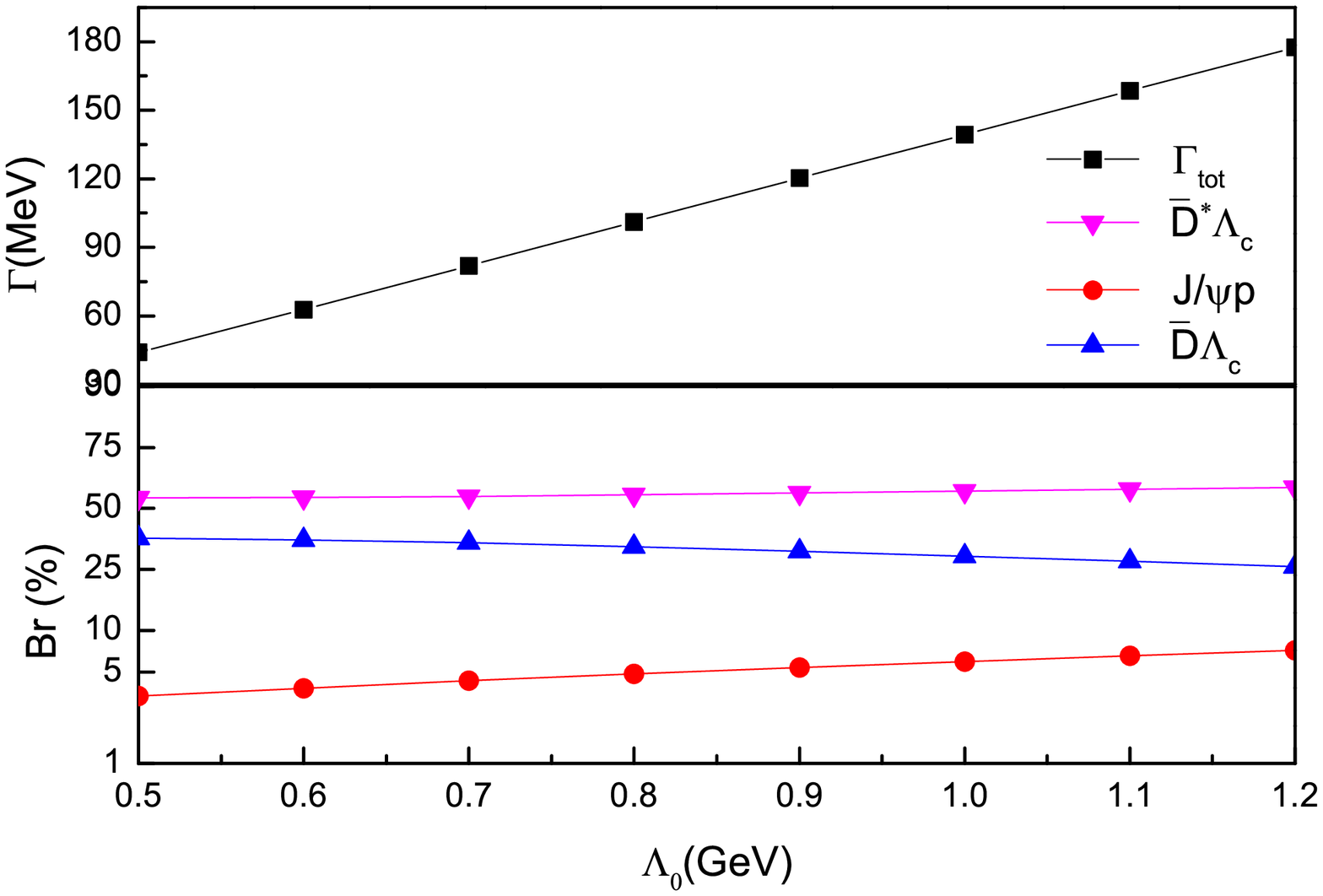} \hfill	
        \includegraphics[width=0.49\textwidth]{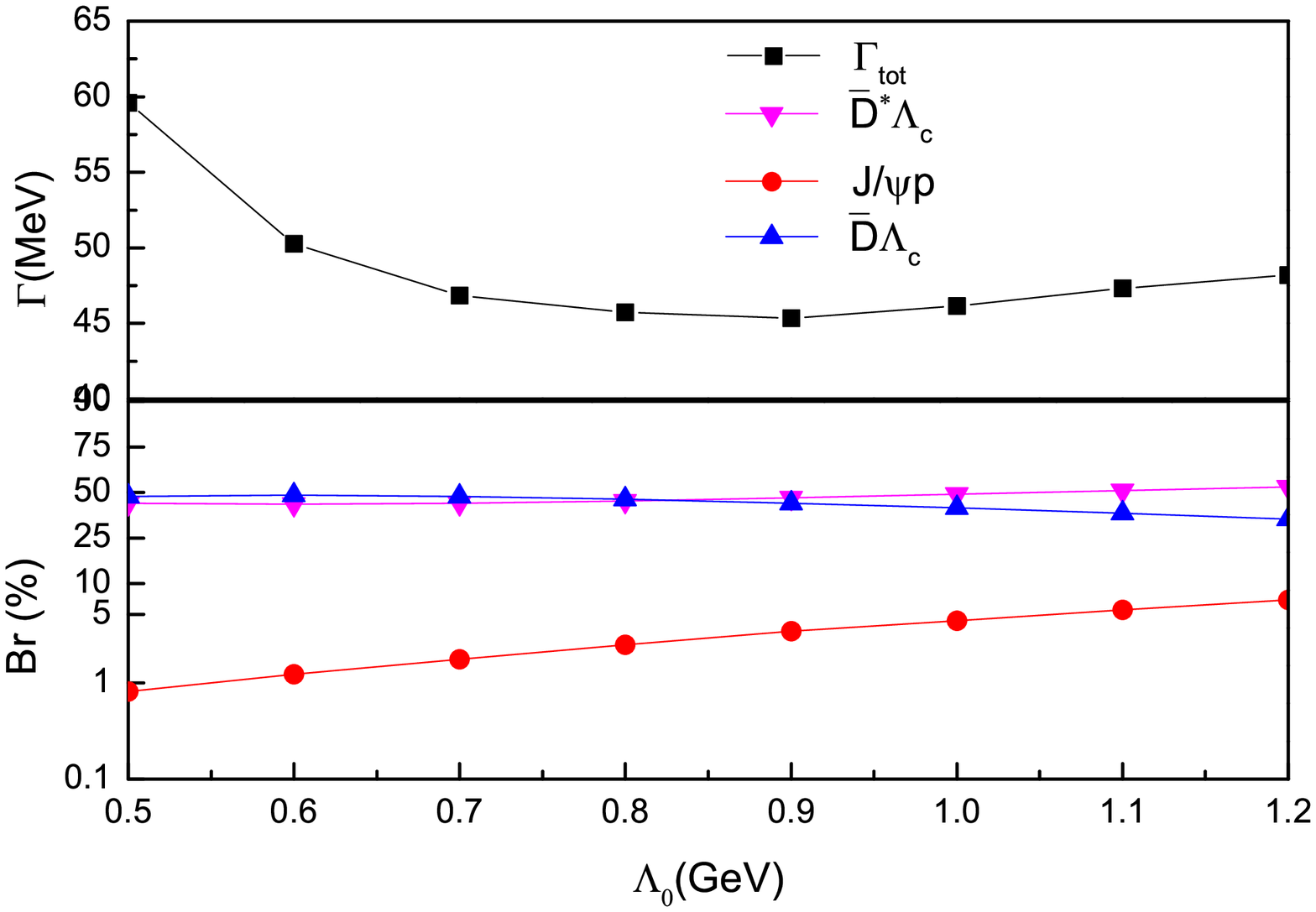}
	\caption{\label{figure:comparison2}Dependence of the $P_c(4450)$ total
width and branching fractions of $\bar
D^*\Lambda_c$, $\bar D\Lambda_c$ and $J/\psi p$ on the cutoff $\Lambda_0$ in
different scenarios for the $P_c(4380)$: (a) $S$-wave $\bar D^*\Sigma_c$
molecule with
$J^P={\frac32}^-$; (b) $P$-wave $\bar D^*\Sigma_c$ molecule with
$J^P={\frac52}^+$. Here
$\Lambda_1$ is fixed at 2.0~GeV.
}
\end{figure}

\begin{figure}[!t]
	\centering
        \includegraphics[width=0.49\textwidth]{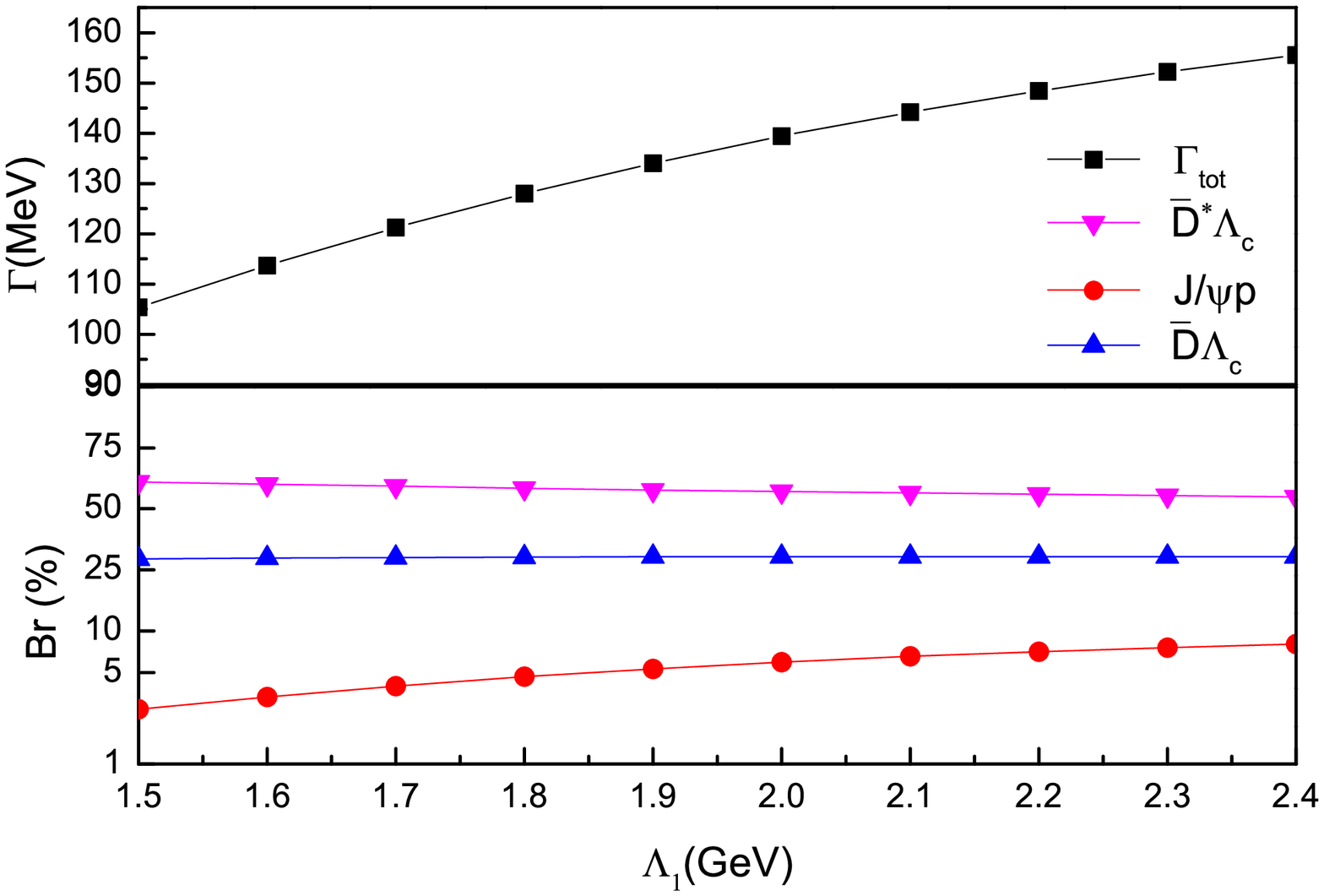}\hfill
        \includegraphics[width=0.49\textwidth]{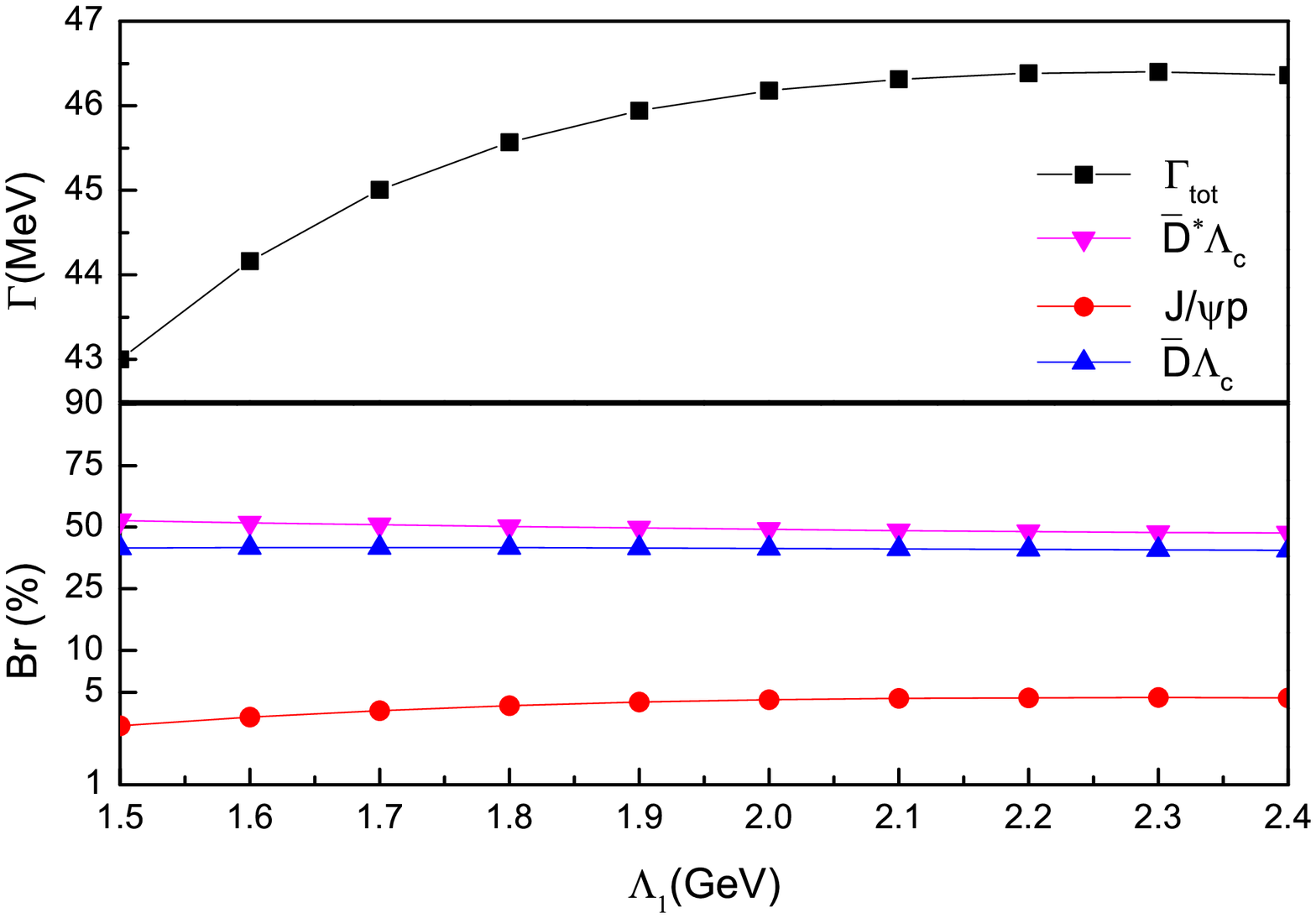}
	\caption{\label{figure:comparison2s}Dependence of the $P_c(4450)$ total
width and branching fractions of $\bar
D^*\Lambda_c$, $\bar D\Lambda_c$ and $J/\psi p$ on the cutoff $\Lambda_1$ in
different scenarios for the $P_c(4380)$: (a) $S$-wave $\bar D^*\Sigma_c$
molecule with
$J^P={\frac32}^-$; (b) $P$-wave $\bar D^*\Sigma_c$ molecule with
$J^P={\frac52}^+$. Here
$\Lambda_0$ is fixed at 1.0~GeV.
 }
\end{figure}

To estimate the partial widths of the $J^P={\frac52}^+$ $P_c(4450)$
state, we use the effective Lagrangian for the $P$-wave interaction among
$P_c(4450)$, $\bar{D}^*$ and $\Sigma_c$ given by~\cite{Zou:2002yy}
\begin{equation}
\Lag_{\bar{D}^* \Sigma_c P_c} = g_{\bar{D}^* \Sigma_c P_c}^{} \left(-g^{\nu \alpha} + \frac{p^{\nu} p^{\alpha}}{p^2}\right) \left(\partial_{\alpha} \bar{\Sigma}_c \bar{D}^{* \mu} - \bar{\Sigma}_c \partial_{\alpha} \bar{D}^{* \mu}\right) P_{\mathrm{c} \mu \nu} + H.c. ,
\label{eq:52vertex*}
\end{equation}
with $p$ the momentum of the $P_c$ state. In analogy
with the $S$-wave interactions described by Eq.~(\ref{eq:vertex0}), the coupling
constant $g_{\bar{D}^* \Sigma_c P_c}^{}$ may be obtained from the
compositeness condition~\cite{Weinberg:1962hj,Aceti:2012dd}. However, being in
a $P$-wave, the obtained coupling strength relies much more on the cutoff
$\Lambda_0$. Thus, we can only make a rough estimate for the widths in
this case. The corresponding numerical results obtained with $\Lambda_0=1.0\
\mathrm{GeV}$ and $\Lambda_1=2.0\ \mathrm{GeV}$ are listed in
Table~\ref{table:total}. The total width of the ${\frac52}^+$ $\bar{D}^*
\Sigma_c$ molecule is about $46\ \mathrm{MeV}$ with these cutoff values. The
dependence on the cutoffs $\Lambda_0$ and $\Lambda_1$ in both $\frac52^+$ and
$\frac32^-$ $\bar D^*\Sigma_c$ scenarios are presented in
Figs.~\ref{figure:comparison2} and~\ref{figure:comparison2s}, respectively.
%
%---------
%---------

%---------
%---------

From the curves of branch ratios, we see that despite of the sizeable cutoff
dependence of the total decay widths, the branching fractions are rather
insensitive to the cutoff values. The decay behaviors in
both the ${\frac52}^+$ and $\frac32^-$ $\bar D^*\Sigma_c$ molecular scenarios
are similar to each other except for two points: the total decay width and the
branch ratio of the $\bar D \Lambda_c$ channel. One sees that within the chosen
cutoff ranges the total width of the $\frac52^+$ $\bar D^*\Sigma_c$ molecule is
almost always much smaller than that in the $\frac32^-$ case, and the former is
in much better agreement with the width reported by the LHCb Collaboration for
the $P_c(4450)$, $(39\pm 5\pm 19)\ \mathrm{MeV}$. From this point of view, the
$\bar{D}^* \Sigma_c$ molecule with $J^P={\frac52}^+$ seems to be a more
favorable assignment for the $P_c(4450)$. The other difference is that the
branching fraction of the $\bar D \Lambda_c$ channel is comparable with that of
the $\bar D^*\Lambda_c$ channel in $J^P={\frac52}^+$ picture, while it is much
smaller in the $J^P={\frac32}^-$ picture. Note that the partial width of the
$\bar D \Lambda_c$ channel decreases with increasing $\Lambda_0$, which leads
to decreasing behavior of the total width of the ${\frac52}^+$ $P_c(4450)$ for
$\Lambda_0\in[0.5,\,0.9] \ \mathrm{GeV}$ with $\Lambda_1=2.0$~GeV.

% \newpage
\section{Production of $P_c$ states in photo- and pion-induced reactions}
\label{sec:production}

\begin{figure}[tb]
\begin{center}
\includegraphics[width=0.42\textwidth]{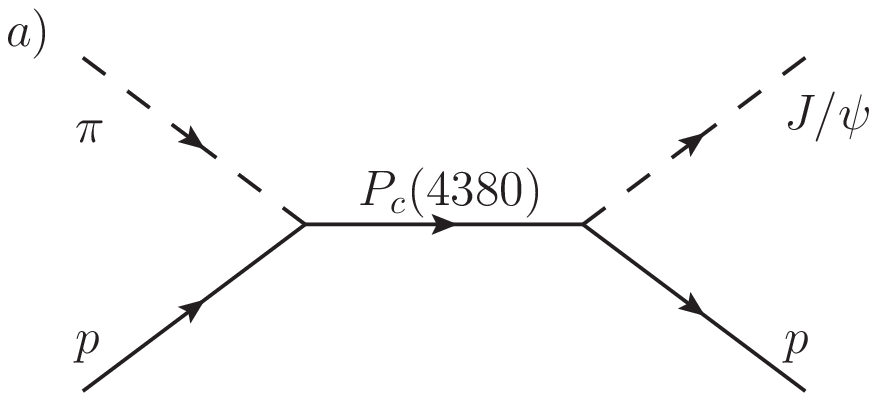}
\includegraphics[width=0.42\textwidth]{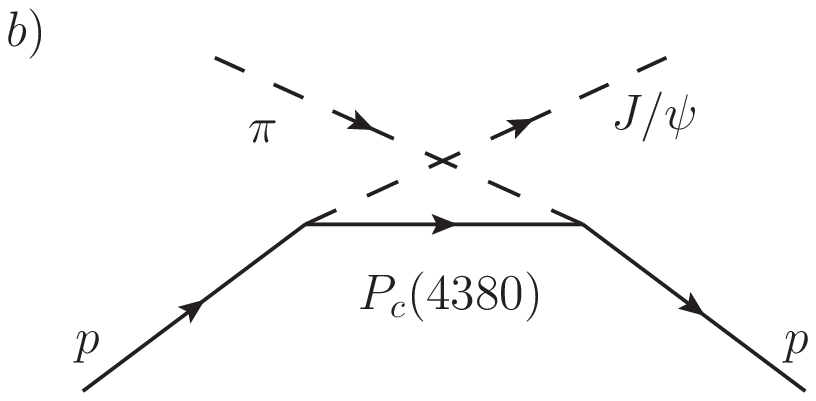}\\ \ \\
\includegraphics[width=0.42\textwidth]{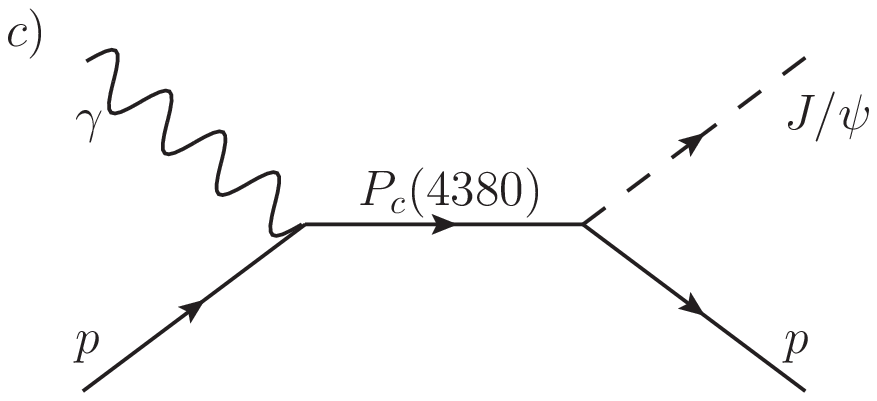}
\includegraphics[width=0.42\textwidth]{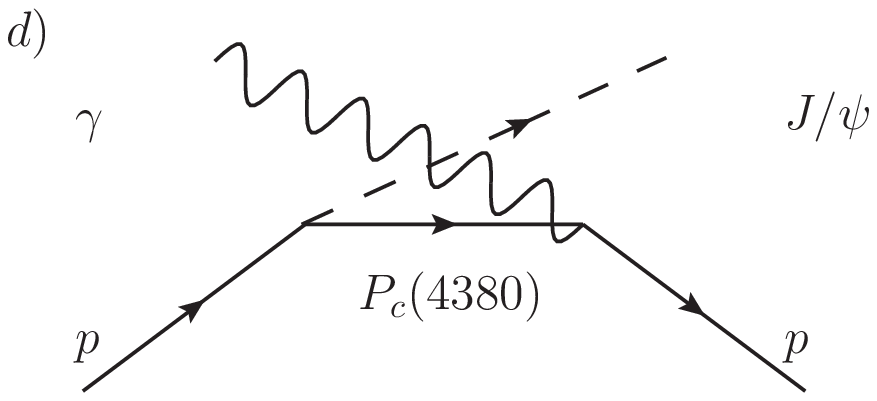}
\caption{The $s$ and $u$-channel reactions with the $P_c(4380)$ as the
intermediate state, where $P_c(4380)$ is treated as the $\bar{D} \Sigma_c^*$
molecule with spin-parity ${\frac32}^{-}$.
a)\&b) The $s$ and $u$-channel contribution of the scattering process $p \pi \to
J/\psi p $.
c)\&d) The $s$ and $u$-channel contribution of the scattering process $p \gamma
\to J/\psi p $.
\label{Fig:tree}}
\end{center}
\end{figure}

In order to further pin down the nature of the $P_c$ states, it would be very
useful to study them through various two-body scattering
processes~\cite{Wang:2015jsa,Lu:2015fva,Kim:2016cxr,Huang:2013mua}. In
particular, this is extremely important so as to distinguish the resonance
scenario from the kinematical
singularities~\cite{Guo:2015umn,Liu:2015fea,Guo:2016bkl,Bayar:2016ftu}. With
the formalism given in the previous section, we can also estimate the total
crossing sections for some common scattering reactions with the $P_c(4380)$ as
the intermediate state, for example the $\gamma p$ and $\pi p$ collisions with
the $J/\psi p$ as the final state, shown in Fig.~\ref{Fig:tree}. Note that the
contribution from the $u$-channel exchange of the $P_c(4380)$ is negligible
compared to the $s$-channel one since the intermediate $P_c$ in
$u$-channel processes will be highly off-shell. Therefore, only the $s$-channel
contribution is included in our calculation. It is similar for the $P_c(4450)$
exchange.
%---------
%---------

% Besides that, the $p \pi P_c$ and $p J/\psi P_c$($p \gamma P_c$) vertices are
% $P B B^*$ and $V B B^*$ type, respectively. But they are different from the
% vertices described by Eq.~(\ref{eq:PBB*}) and Eq.~(\ref{eq:VBB*}) since the
% $B^*$ is a spin-parity ${\frac32}^-$ baryon in these tree diagrams.

We consider the possibility of $J^P=\frac32^-$ for the $P_c(4380)$. The $p
\pi P_c$ and $p J/\psi P_c$($p \gamma P_c$) vertices should be dominated by
$D$-wave and $S$-wave, respectively. The effective Lagrangians for these two
kinds of vertices are given by~\cite{Lu:2015fva}
\begin{eqnarray}
\Lag_{P B P_c} &=& g_{P B P_c}^{} \bar{P}_c^{ \mu}
\gamma_5 \gamma_{\nu} \partial^{\nu} \partial_{\mu} P B + \text{H.c.},
\label{eq:PBB1*}\\
\Lag_{V B P_c} &=& -i g_{V B P_c}^{} \bar{P}_c^{\mu}
\gamma^{\nu}  B F_{\mu\nu} + \text{H.c.},
\label{eq:VBB1*}
\end{eqnarray}
where $P$ and $B$ are the fields for the pion and proton, respectively,
$F_{\mu\nu}=\partial_\mu V_\nu - \partial_\nu V_\mu$ with $V$ the field for the
photon or $J/\psi$.

\begin{figure}[htbp]
\begin{center}
\includegraphics[width=0.50\textwidth]{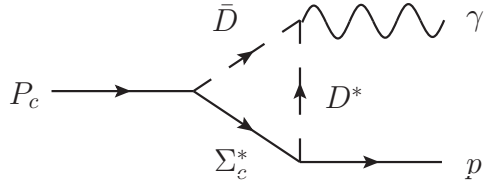}
\caption{The decay of the $P_c$ state into $p \gamma$.
\label{Fig:gamma}}
\end{center}
\end{figure}

To study these two reactions in Fig.~\ref{Fig:tree}, we need to get the related
coupling constants $g_{p J/\psi P_c}$, $g_{p \pi P_c}$ and $g_{p \gamma P_c}$.
These coupling constants can be deduced form the partial widths of the $P_c$
state decaying into $J/\psi p$, $\pi p$ and $p \gamma$, respectively, which
are estimated by calculating the triangle diagrams as in
Section~\ref{sec:results}, where the partial widths for the $p\pi$ and $p
J/\psi$ have been calculated. For the $p \gamma$ channel, the needed
triangle diagram is shown in Fig.~\ref{Fig:gamma}.
%---------
%---------
The values of these couplings and the partial widths of the $P_c(4380)$
as a $J^P = {\frac32}^-$ $\bar{D} \Sigma_c^*$ molecule that we will use in
estimating the cross sections are listed in Table~\ref{table:amplitude}.
%---------
\begin{table}[htbp]
\centering
\caption{\label{table:amplitude}Partial widths and couplings of $P_c(4380)$ as a
$J^P={\frac32}^-$ $\bar D \Sigma_c^*$ molecule into $J/\psi p$, $\pi p$ and $p
\gamma$ final states calculated with $\Lambda_0=1.0\ \mathrm{GeV}$,
$\Lambda_1=2.0\ \mathrm{GeV}$. }
% \scalebox{1.2}{
\begin{tabular}{l|*{2}{c}}
\Xhline{1pt}
\thead{Modes} & \thead{Widths($\mathrm{MeV}$)} & \thead{Couplings} \\
\Xhline{0.8pt}
$J/\psi p$ 	 & 3.8     & 0.11\ ($\mathrm{GeV}^{-1}$) 	\\
$\pi p$ 	 & 0.06     & 0.0053\ ($\mathrm{GeV}^{-2}$) 	\\
$p \gamma$   & 0.0007  & 0.00097\ ($\mathrm{GeV}^{-1}$) 	\\
\Xhline{1pt}
\end{tabular}
% }
\end{table}
%---------

\begin{figure}[htbp]
	\centering
	\includegraphics[width=0.49\textwidth]{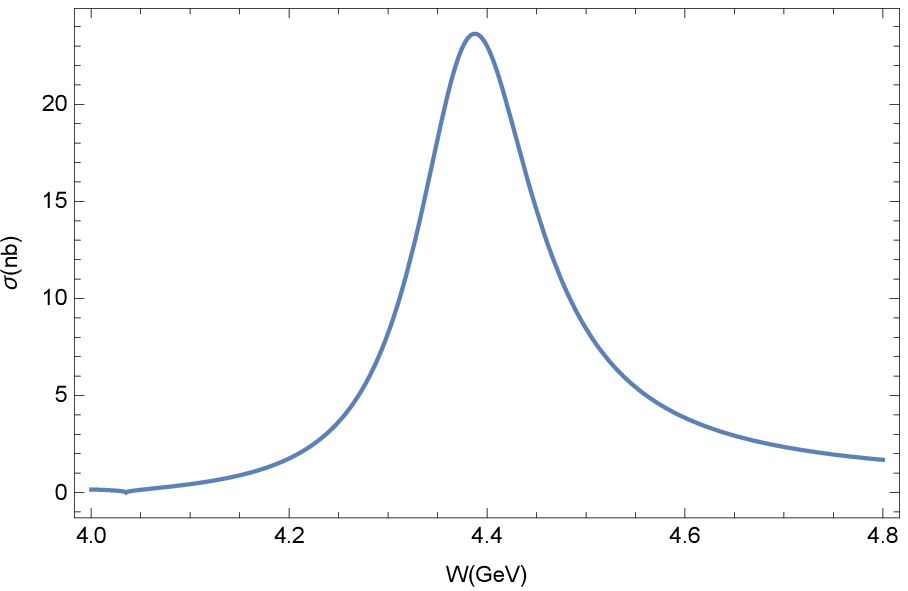}\hfill
	\includegraphics[width=0.49\textwidth]{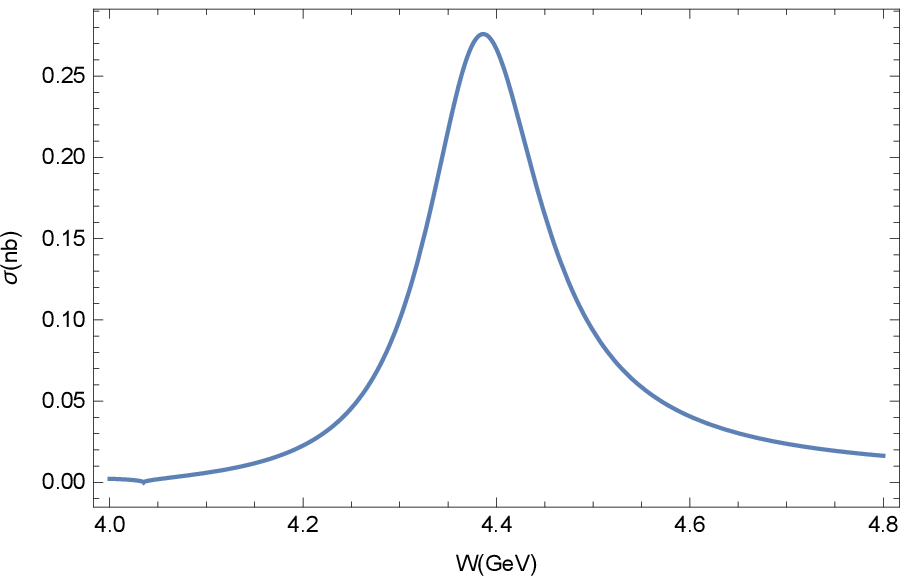}
	\caption{ Dependence of the total cross sections on the
center-of-mass energy $W$ for the reactions (a) $p \pi \to J/\psi
p $ and (b) $p \gamma \to J/\psi p $.
\label{Fig:sigma}}
\end{figure}

With the above Lagrangians and coupling constants,
the cross sections can be estimated immediately by computing the tree diagrams
shown in Fig.~\ref{Fig:gamma}. The numerical results are given in
Fig.~\ref{Fig:sigma}.  The clear peak structure in the cross sections is due to
the $s$-channel exchange of the $P_c(4380)$ resonance in the reactions $p \pi
\to J/\psi p $ and $p \gamma \to J/\psi p $.

%---------
%---------

The numerical calculation suggests that the probability for the $\pi p$ to
$J/\psi p $ through $P_c(4380)$ molecule is two orders of magnitude larger
than that in the $\gamma p$ collisions. This can be easily understood as the
photo-production cross section is roughly suppressed by a factor of the fine
structure constant $\alpha=1/137$.
% It coincides with the results on the
% spin-parity ${\frac32}^-$ $P_c$ state presented in
% Refs.~\cite{Wang:2015jsa,Lu:2015fva,Kim:2016cxr,Huang:2013mua}.
The cross sections of the processes $\pi p\to J/\psi p $ and $\gamma p\to J/\psi
p $ reactions have been calculated in Refs.~\cite{Lu:2015fva,Kim:2016cxr} and
\cite{Wang:2015jsa,Huang:2013mua}, respectively, by assuming that
$J/\psi p$ and $\pi p$ channels account for some specific percentages of the
$P_c(4380)$ total width. In Refs.~\cite{Wang:2015jsa,Huang:2013mua}, the
coupling constant $g_{p \gamma P_c}$ was estimated using the
vector-meson-dominance model by assuming that the photon couples through
intermediate vector mesons ($J/\psi$ in Ref.~\cite{Wang:2015jsa} and $\omega$,
$\rho$ and $J/\psi$ in Ref.~\cite{Huang:2013mua}). The coupling
constant $g_{pJ/\psi P_c}$ used in Ref.~\cite{Huang:2013mua} is deduced from
the partial width of $P_c \to pJ/\psi$ predicted in Ref.~\cite{Wu:2010jy}
prior to the $P_c$ discovery. The estimation in Ref.~\cite{Huang:2013mua} for
the total cross section of $\gamma p\to J/\psi p $ is in the order of magnitude
of $0.1$ to $1$~$\mathrm{nb}$, which in line with the result in
Ref.~\cite{Wang:2015jsa} with the $g_{p J/\psi P_c}$ obtained by assuming a
$5\%$  branching fraction for the $P_c\to J/\psi p$.
Refs.~\cite{Lu:2015fva} and \cite{Kim:2016cxr} obtained different cross
sections for the $\pi p\to J/\psi p $ by using different decay branching
ratios for the $J/\psi p$ and $p \pi$ channels. The former claims the cross
section of $p \pi \to J/\psi p $ is of order $1\
\mathrm{\mu b}$ by assuming the branching ratios are $10\%$ and  $1\%$ for
the $J/\psi p$ and $p \pi$ channels, respectively, while the result is
at the level of $1\ \mathrm{nb}$ by assuming branching fractions of $5\%$
and $10^{-5}$ for the $J/\psi p$ and $p \pi$ channels, respectively, in
Ref.~\cite{Kim:2016cxr}. However, in our work, we have obtained the partial
widths of $P_c(4380)$ into these channels in the preceding calculations. Based
on the results obtained using $\Lambda_0=1.0\ \mathrm{GeV}$,
$\Lambda_1=2.0\ \mathrm{GeV}$,  the total decay width of
$P_c(4380)$ is $144\ \mathrm{MeV}$, and the branching ratios of $J/\psi p$ and
$p \pi$ channels are around $3\%$ and $0.04\%$, respectively. Apparently, the
cross sections of these tree diagrams are determined by these values. The
parameters obtained from our calculations are different from the assumptions in
these literature, and thus we obtain different cross sections.

In addition, another interesting conclusion can be deduced from our numerical
results. As shown in Table~\ref{table:total}, the $\bar D^*\Lambda_c$ and
$\bar D\Lambda_c$ channels should be the dominant channels for both
the $P_c(4380)$ as a $J^P = {\frac32}^-$ $\bar{D} \Sigma_c^*$ molecule
and the $P_c(4450)$ as a $J^P = {\frac52}^+$ $\bar{D}^* \Sigma_c$
molecule. Their partial widths are much larger than that of the $J/\psi p$.
This means that the cross sections of the processes $\pi p\to \bar D^*\Lambda_c
$ and $\pi p\to \bar D\Lambda_c$ through $P_c(4380)$ and $P_c(4450)$ must be
much larger than the reaction $\pi p\to J/\psi p $. It is consistent with the
claims in Refs.~\cite{Kim:2016imp,Kim:2016cxr}. In conclusion, from the point
of view of cross sections, it should be easier to search for the pentaquark
states with hidden-charm in the $\bar D^*\Lambda_c$ and $\bar D\Lambda_c$
production than the $J/\psi p$ production.
% Furthermore, these cross section
% distributions could be measured to check our results in future experiments.

\section{Summary} \label{sec:summary}

An interesting property of the two $P_c$ structures reported by the LHCb
Collaboration in 2015 is that they are located just below the $\bar{D}
\Sigma_c^*(2520)$ and $\bar{D}^* \Sigma_c(2455)$ thresholds, respectively.
Inspired by this property, the two $P_c$ states
were proposed to be either $\bar{D} \Sigma_c^*$ or $\bar{D}^* \Sigma_c$
$S$-wave bound states of spin-parity $J^P={\frac32}^-$. We estimated the decay
behaviors of such two types of hadronic molecules in this paper. With branching
ratios of ten possible decay channels calculated, it is found that the two types
of hadronic molecules have distinguishable decay patterns. While the $\bar{D}
\Sigma_c^*$ molecule decays dominantly to the $\bar{D}^* \Lambda_c$ channel with
a branching ratio by two orders of magnitude larger than that to the
$\bar{D}\Lambda_c$, the $\bar{D}^* \Sigma_c$ molecule decays into these two
channels with a difference of less than a factor of 2. Our results show that the
total decay width of $P_c(4380)$ as a ${\frac32}^-$ $\bar{D}
\Sigma_c^*$ molecule is about a factor of 2 larger than the corresponding value
for the $\bar{D}^* \Sigma_c$ molecule. It seems to suggest that the assignment
of $\bar{D} \Sigma_c^*$ molecule for the $P_c(4380)$ is more favorable than the
$\bar{D}^* \Sigma_c$ molecule. The results for the $P_c(4450)$ indicate that the
$P_c(4450)$ is more likely a $J^P = {\frac52}^+$ $\bar{D}^* \Sigma_c$ $P$-wave
molecule than an $\frac32^-$ $\bar{D}^* \Sigma_c$ $S$-wave molecule. In order
to further pin down the nature of the $P_c$ states, it would be very useful to
study them through various two-body scattering processes.
Based on the partial decay widths of the $P_c(4380)$, we estimated the cross
sections for the reactions $\gamma p \to J/\psi p $ and $ \pi p\to J/\psi p $
through exchanging the $P_c(4380)$ state in the $s$-channel. The peak values
are at the level of 0.2~nb and 20~nb, respectively. The corresponding
productions rates for reactions into $\bar D^*\Lambda_c$ and/or $\bar
D\Lambda_c$ would be larger by orders of magnitude. The forthcoming $\gamma p$
experiment at JLAB and $\pi p$ experiment at JPARC should be able to provide
valuable information towards revealing the nature of these $P_c$
structures.

\bigskip

\section*{Acknowledgments}

We thank Yu Lu for helps on computing program. This project is supported by
NSFC under Grant No.~11261130311 (CRC110 cofunded by DFG and NSFC) and Grant
No.~11647601, and by the Thousand Talents Plan for Young Professionals, and by
the CAS Key Research Program of Frontier Sciences  under Grant
No.~QYZDB-SSW-SYS013.

% \newpage
\bigskip

\begin{appendix}

\section*{Appendix: decay amplitudes}

\begin{figure}[htpb]
\begin{center}
\includegraphics[width=0.31\textwidth]{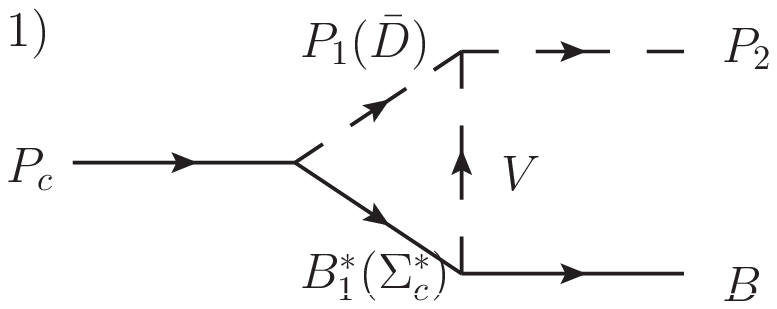}
\includegraphics[width=0.31\textwidth]{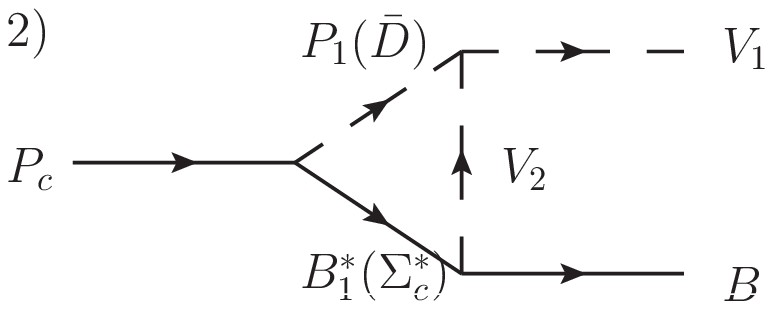}
\includegraphics[width=0.31\textwidth]{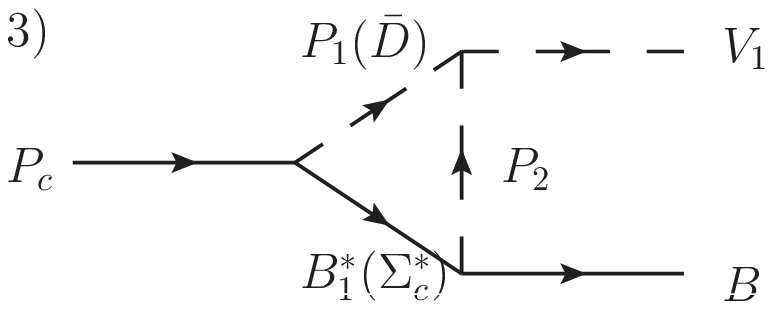}\\ \ \\
\includegraphics[width=0.31\textwidth]{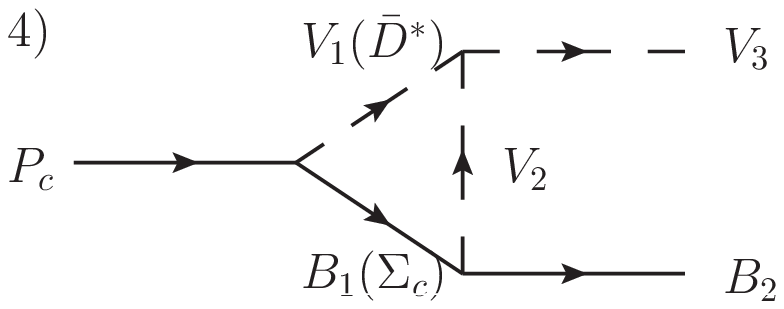}
\includegraphics[width=0.31\textwidth]{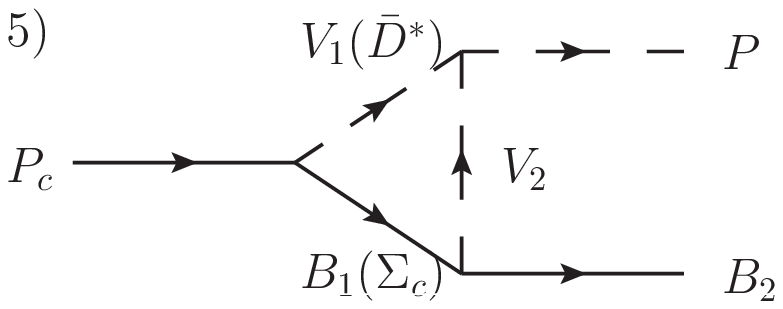}
\includegraphics[width=0.31\textwidth]{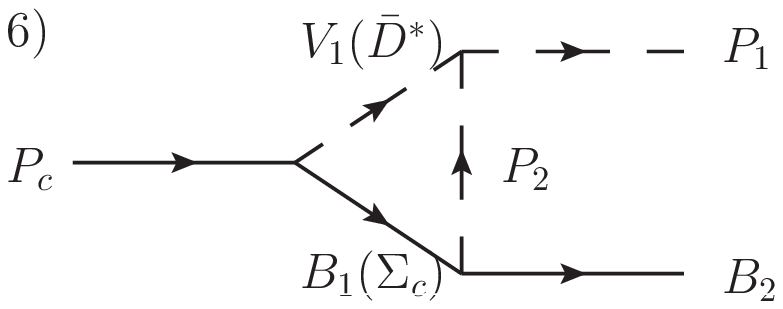}\\ \ \\
\includegraphics[width=0.31\textwidth]{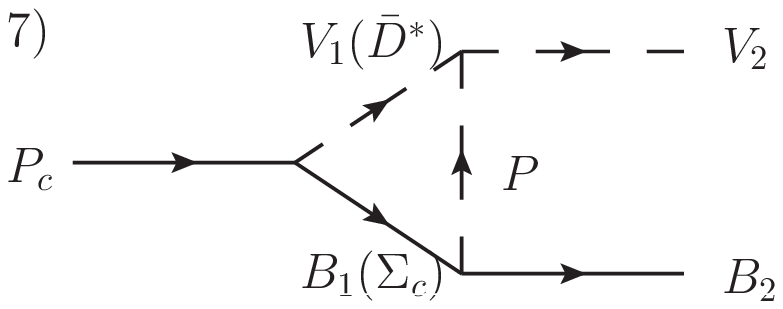}
\caption{Various kinds of triangle diagrams for the two-body decays of the
$P_c$ state.
$1$) The $\bar D\Lambda_c$ channel with $\rho$ exchange, $\pi N$ channel with $D^*$ exchange, $\eta_c p$ channel with $D^*$ exchange and $\bar D\Sigma_c$ channel with $\rho$ exchange for the $\bar D\Sigma^*_c$ hadronic molecule of $P_c(4380)$.
$2$) $\bar D^*\Lambda_c$ channel with $\rho$ exchange, $J/\psi p$ channel with $D^*$ exchange, $\rho N$ channel with $D^*$ exchange and $\omega p$ channel with $D^*$ exchange for the $\bar D\Sigma^*_c$ hadronic molecule of $P_c(4380)$.
$3$) $\bar D^*\Lambda_c$ channel with $\pi$ exchange, $J/\psi p$ channel with $D$ exchange, $\rho N$ channel with $D$ exchange and $\omega p$ channel with $D$ exchange for the $\bar D\Sigma^*_c$ hadronic molecule of $P_c(4380)$.
$4$) $\bar D^*\Lambda_c$ channel with $\rho$ exchange, $J/\psi p$ channel with $D^*$ exchange, $\rho N$ channel with $D^*$ exchange and $\omega p$ channel with $D^*$ exchange for the $\bar{D}^* \Sigma_c$ hadronic molecules of $P_c(4380)$ and $P_c(4450)$.
$5$) The $\bar D\Lambda_c$ channel with $\rho$ exchange, $\pi N$ channel with $D^*$ exchange, $\eta_c p$ channel with $D^*$ exchange and $\bar D\Sigma_c$ channel with $\rho$ exchange for the $\bar{D}^* \Sigma_c$ hadronic molecules of $P_c(4380)$ and $P_c(4450)$.
$6$) The $\bar D\Lambda_c$ channel with $\pi$ exchange, $\pi N$ channel with $D$ exchange, $\eta_c p$ channel with $D$ exchange and $\bar D\Sigma_c$ channel with $\pi$ exchange for the $\bar{D}^* \Sigma_c$ hadronic molecules of $P_c(4380)$ and $P_c(4450)$.
$7$) $\bar D^*\Lambda_c$ channel with $\pi$ exchange, $J/\psi p$ channel with $D$ exchange, $\rho N$ channel with $D$ exchange and $\omega p$ channel with $D$ exchange for the $\bar{D}^* \Sigma_c$ hadronic molecules of $P_c(4380)$ and $P_c(4450)$.
\label{Fig:kinds}}
\end{center}
\end{figure}

This appendix collects together all the formulae that are used in the
calculations of the scattering amplitudes in our work. Except for the
$\chi_{c0} p$ final states described by diagram $e)$, $e^{\prime})$ in
Fig.~\ref{Fig: mechanism},~\ref{Fig: mechanism1}, and the $\bar D\Sigma^*_c$
channels shown in Fig.~\ref{Fig:4450}, the other two-body decays for the two
$P_c$ states can be classified into the categories shown in
Fig.~\ref{Fig:kinds}.

The corresponding amplitudes can be written as
\begin{eqnarray}
\mM_1 &=& g_{\bar{D} \Sigma_c^* P_c}^{} g_{V P_1 P_2}^{} g_{V B B_1^*}^{} \int_{-\infty}^{\infty} \frac{{\rm d}^4 k}{(2
\pi)^4} \Biggl\{ \tilde\Phi\left[ \frac{m_{\bar{D}}}{m_{\bar{D}} + m_{\Sigma_c^*}} p_0 - (p_1 + k) \right] \frac{\Lambda_1^4}{(m_V^2 - k^2)^2 + \Lambda_1^4} \notag \\
&& \mbox{} \qquad \bar{u}(p_1, s_B) \gamma^{\alpha} \gamma_5 \frac1{k^2-m_V^2}\left[k_\mu \left(g_{\alpha \beta} - \frac{k_{\alpha} k_{\beta}}{m_V^2}\right) - k_\alpha \left(g_{\mu \beta} - \frac{k_{\mu} k_{\beta}}{m_V^2}\right) \right] \notag \\
&& \mbox{} \frac{\slashed{p_1} + \slashed{k} + m_{\Sigma_c^*}}{(p_1+k)^2-m_{\Sigma_c^*}^2+i m_{\Sigma_c^*} \Gamma_{\Sigma_c^*}} \biggl[-g^{\mu \nu}+\frac13 \gamma^\mu \gamma^\nu +\frac1{3 m_{\Sigma_c^*}} (\gamma^\mu (p_1+k)^\nu-\gamma^\nu (p_1+k)^\mu) + \notag \\
 && \mbox{} \frac2{3 m_{\Sigma_c^*}^2} (p_1+k)^\mu (p_1+k)^\nu \biggr] \left[-2 (p_0-p_1)+k\right]^\beta \frac1{(p_0-p_1-k)^2-m_{P_1}^2}u_{\nu}(p_0, s_{P_c}) \Biggr\} ,
\end{eqnarray}

\begin{eqnarray}
\mM_2 &=& g_{\bar{D} \Sigma_c^* P_c}^{} g_{V_1 V_2 P_1}^{} g_{V_2 B B_1^*}^{} \int_{-\infty}^{\infty} \frac{{\rm d}^4 k}{(2
\pi)^4} \Biggl\{ \tilde\Phi\left[ \frac{m_{\bar{D}}}{m_{\bar{D}} + m_{\Sigma_c^*}} p_0 - (p_1 + k) \right] \frac{\Lambda_1^4}{(m_{V_2}^2 - k^2)^2 + \Lambda_1^4} \notag \\
&& \mbox{} \qquad \bar{u}(p_1, s_B) \gamma^{\alpha} \gamma_5 \frac1{k^2-m_{V_2}^2}\left[k_\mu \left(g_{\alpha \beta} - \frac{k_{\alpha} k_{\beta}}{m_{V_2}^2}\right) - k_\alpha \left(g_{\mu \beta} - \frac{k_{\mu} k_{\beta}}{m_{V_2}^2}\right) \right] \notag \\
&& \mbox{} \frac{\slashed{p_1} + \slashed{k} + m_{\Sigma_c^*}}{(p_1+k)^2-m_{\Sigma_c^*}^2+i m_{\Sigma_c^*} \Gamma_{\Sigma_c^*}} \biggl[-g^{\mu \nu}+\frac13 \gamma^\mu \gamma^\nu +\frac1{3 m_{\Sigma_c^*}} (\gamma^\mu (p_1+k)^\nu-\gamma^\nu (p_1+k)^\mu) + \notag \\
 && \mbox{} \frac2{3 m_{\Sigma_c^*}^2} (p_1+k)^\mu (p_1+k)^\nu \biggr] \varepsilon^{\rho \sigma \lambda \beta} \left[-(p_0-p_1)_\rho \epsilon^*_{\sigma}(p_0-p_1, s_{V_1}) k_{\lambda}\right] \frac1{(p_0-p_1-k)^2-m_{P_1}^2} \notag \\
 && \mbox{} \qquad\qquad u_{\nu}(p_0, s_{P_c}) \Biggr\}, \\
 \mM_3 &=& -g_{\bar{D} \Sigma_c^* P_c}^{} g_{V_1 P_1 P_2}^{} g_{P_2 B B_1^*}^{} \int_{-\infty}^{\infty} \frac{{\rm d}^4 k}{(2
\pi)^4} \Biggl\{ \tilde\Phi\left[ \frac{m_{\bar{D}}}{m_{\bar{D}} + m_{\Sigma_c^*}} p_0 - (p_1 + k) \right] \frac{\Lambda_1^4}{(m_{P_2}^2 - k^2)^2 + \Lambda_1^4} \notag \\
&& \mbox{} \qquad \bar{u}(p_1, s_B) k_\mu \frac{\slashed{p_1} + \slashed{k} + m_{\Sigma_c^*}}{(p_1+k)^2-m_{\Sigma_c^*}^2+i m_{\Sigma_c^*} \Gamma_{\Sigma_c^*}} \biggl[-g^{\mu \nu}+\frac13 \gamma^\mu \gamma^\nu +\frac1{3 m_{\Sigma_c^*}} \bigl(\gamma^\mu (p_1+k)^\nu- \notag \\
&& \mbox{} \qquad \gamma^\nu (p_1+k)^\mu \bigr) + \frac2{3 m_{\Sigma_c^*}^2} (p_1+k)^\mu (p_1+k)^\nu \biggr] \left[2k-(p_0-p_1)\right]^\beta \epsilon^*_{\beta}(p_0-p_1, s_{V_1}) \notag \\
&& \mbox{} \qquad\qquad \frac1{k^2-m_{P_2}^2} \frac1{(p_0-p_1-k)^2-m_{P_1}^2}u_{\nu}(p_0, s_{P_c}) \Biggr\},
\end{eqnarray}

\begin{eqnarray}
\mM_4 &=& -i g_{\bar{D}^* \Sigma_c P_c}^{} g_{V_1 V_2 V_3}^{} g_{V_2 B_1 B_2}^{} \int_{-\infty}^{\infty} \frac{{\rm d}^4 k}{(2
\pi)^4} \Biggl\{ \tilde\Phi\left[ \frac{m_{\bar{D}^*}}{m_{\bar{D}^*} + m_{\Sigma_c}} p_0 - (p_1 + k) \right] \frac{\Lambda_1^4}{(m_{V_2}^2 - k^2)^2 + \Lambda_1^4} \notag \\
&& \mbox{} \qquad \bar{u}(p_1, s_{B_2}) \gamma^\alpha \frac1{k^2-m_{V_2}^2} \left(g_{\alpha \beta} - \frac{k_{\alpha} k_{\beta}}{m_{V_2}^2}\right) \frac{\slashed{p_1} + \slashed{k} + m_{\Sigma_c}}{(p_1+k)^2-m_{\Sigma_c}^2} \biggl[\left(2k-(p_0-p_1)\right)^\mu g^{\beta \sigma} + \notag \\
&& \mbox{} \qquad\qquad \left(-(p_0-p_1)-k\right)^\sigma g^{\mu \beta} + \left(2(p_0-p_1) -k\right)^\beta g^{\mu \sigma} \biggr] \frac1{(p_0-p_1-k)^2-m_{V_1}^2} \notag \\
&& \mbox{} \left(g_{\sigma \nu} - \frac{(p_0-p_1-k)_{\sigma}
(p_0-p_1-k)_{\nu}}{m_{V_1}^2}\right) \epsilon^*_{\mu}(p_0-p_1, s_{V_3})
u^{\nu}(p_0, s_{P_c}) \Biggr\},
\end{eqnarray}

\begin{eqnarray}
\mM_5 &=& i g_{\bar{D}^* \Sigma_c P_c}^{} g_{V_1 V_2 P}^{} g_{V_2 B_1 B_2}^{} \int_{-\infty}^{\infty} \frac{{\rm d}^4 k}{(2
\pi)^4} \Biggl\{ \tilde\Phi\left[ \frac{m_{\bar{D}^*}}{m_{\bar{D}^*} + m_{\Sigma_c}} p_0 - (p_1 + k) \right] \frac{\Lambda_1^4}{(m_{V_2}^2 - k^2)^2 + \Lambda_1^4} \notag \\
&& \mbox{} \qquad \bar{u}(p_1, s_{B_2}) \gamma^\mu \frac{\slashed{p_1} + \slashed{k} + m_{\Sigma_c}}{(p_1+k)^2-m_{\Sigma_c}^2} \frac1{k^2-m_{V_2}^2} \left(g_{\mu \alpha} - \frac{k_{\mu} k)_{\alpha}}{m_{V_2}^2}\right) \varepsilon^{\rho \alpha \sigma \beta} \left[k_\rho (p_0-p_1-k)_\sigma \right] \notag \\
&& \mbox{} \qquad \frac1{(p_0-p_1-k)^2-m_{V_1}^2} \left[g_{\beta \nu} -
\frac{(p_0-p_1-k)_{\beta} (p_0-p_1-k)_{\nu}}{m_{V_1}^2}\right] u^{\nu}(p_0,
s_{P_c}) \Biggr\},\\
\mM_6 &=& i g_{\bar{D}^* \Sigma_c P_c}^{} g_{V_1 P_1 P_2}^{} g_{P_2 B_1 B_2}^{} \int_{-\infty}^{\infty} \frac{{\rm d}^4 k}{(2
\pi)^4} \Biggl\{ \tilde\Phi\left[ \frac{m_{\bar{D}^*}}{m_{\bar{D}^*} + m_{\Sigma_c}} p_0 - (p_1 + k) \right] \frac{\Lambda_1^4}{(m_{P_2}^2 - k^2)^2 + \Lambda_1^4} \notag \\
&& \mbox{} \qquad \bar{u}(p_1, s_{B_2}) \gamma^5 \frac{\slashed{p_1} + \slashed{k} + m_{\Sigma_c}}{(p_1+k)^2-m_{\Sigma_c}^2} \frac1{(p_0-p_1-k)^2-m_{V_1}^2} \left[g^{\beta \nu} - \frac{(p_0-p_1-k)^{\beta} (p_0-p_1-k)^{\nu}}{m_{V_1}^2}\right] \notag \\
&& \mbox{} \qquad\qquad \left(-(p_0-p_1)-k\right)_\beta \frac1{k^2-m_{P}^2}
u_{\nu}(p_0, s_{P_c}) \Biggr\}, \\
% \end{eqnarray}
% \begin{eqnarray}
\mM_7 &=& i g_{\bar{D}^* \Sigma_c P_c}^{} g_{V_1 V_2 P}^{} g_{P B_1 B_2}^{} \int_{-\infty}^{\infty} \frac{{\rm d}^4 k}{(2
\pi)^4} \Biggl\{ \tilde\Phi\left[ \frac{m_{\bar{D}^*}}{m_{\bar{D}^*} + m_{\Sigma_c}} p_0 - (p_1 + k) \right] \frac{\Lambda_1^4}{(m_P^2 - k^2)^2 + \Lambda_1^4} \notag \\
&& \mbox{} \qquad \bar{u}(p_1, s_{B_2}) \gamma^5 \frac{\slashed{p_1} + \slashed{k} + m_{\Sigma_c}}{(p_1+k)^2-m_{\Sigma_c}^2} \varepsilon^{\alpha \beta \lambda \rho} \left[(p_0-p_1-k)_\alpha (-(p_0-p_1))_\lambda \epsilon^*_{\rho}(p_0-p_1, s_{V_2}) \right] \notag \\
&& \mbox{} \qquad \frac1{(p_0-p_1-k)^2-m_{V_1}^2} \left[g_{\beta \nu} - \frac{(p_0-p_1-k)_{\beta} (p_0-p_1-k)_{\nu}}{m_{V_1}^2}\right] \notag \\
&& \mbox{} \qquad\qquad \frac1{k^2-m_{P}^2} u^{\nu}(p_0, s_{P_c}) \Biggr\},
\end{eqnarray}

As for the $\chi_{c0} p$ and $\bar D\Sigma^*_c$ channels, the expressions of the amplitudes are given as,

\begin{eqnarray}
\mM_{\chi_{c0} p-D} &=& -i g_{\bar{D} \Sigma_c^* P_c}^{} g_{\bar{D} D \chi_{c0}}^{} g_{D p \Sigma_c^*}^{} \int_{-\infty}^{\infty} \frac{{\rm d}^4 k}{(2
\pi)^4} \Biggl\{ \tilde\Phi\left[ \frac{m_{\bar{D}}}{m_{\bar{D}} + m_{\Sigma_c^*}} p_0 - (p_1 + k) \right] \frac{\Lambda_1^4}{(m_D^2 - k^2)^2 + \Lambda_1^4} \notag \\
&& \mbox{} \bar{u}(p_1, s_p) k_{\mu} \frac{\slashed{p_1} + \slashed{k} + m_{\Sigma_c^*}}{(p_1+k)^2-m_{\Sigma_c^*}^2+i m_{\Sigma_c^*} \Gamma_{\Sigma_c^*}} \biggl[-g^{\mu \nu}+\frac13 \gamma^\mu \gamma^\nu +\frac1{3 m_{\Sigma_c^*}} \bigl(\gamma^\mu (p_1+k)^\nu-\notag \\
&& \mbox{} \gamma^\nu (p_1+k)^\mu \bigr) \frac2{3 m_{\Sigma_c^*}^2} (p_1+k)^\mu (p_1+k)^\nu \biggr] \frac1{k^2-m_{D}^2} \frac1{(p_0-p_1-k)^2-m_{\bar{D}}^2} \notag \\
&& \mbox{} \qquad \qquad u_{\nu}(p_0, s_{P_c}) \Biggr\} ,
\end{eqnarray}

\begin{eqnarray}
\mM_{\chi_{c0} p-D^*} &=& -g_{\bar{D}^* \Sigma_c P_c}^{} g_{\bar{D}^* D^* \chi_{c0}}^{} g_{D^* p \Sigma_c}^{} \int_{-\infty}^{\infty} \frac{{\rm d}^4 k}{(2
\pi)^4} \Biggl\{ \tilde\Phi\left[ \frac{m_{\bar{D}^*}}{m_{\bar{D}^*} + m_{\Sigma_c}} p_0 - (p_1 + k) \right] \frac{\Lambda_1^4}{(m_{D^*}^2 - k^2)^2 + \Lambda_1^4} \notag \\
&& \mbox{} \bar{u}(p_1, s_p) \gamma^{\alpha} \frac1{k^2-m_{D^*}^2}\left(g_{\alpha \beta} - \frac{k_{\alpha} k_{\beta}}{m_{D^*}^2}\right) \frac{\slashed{p_1} + \slashed{k} + m_{\Sigma_c}}{(p_1+k)^2-m_{\Sigma_c}^2} \frac1{(p_0-p_1-k)^2-m_{\bar{D}^*}^2} \notag \\
&& \mbox{} \qquad\left[g^{\beta \nu} - \frac{(p_0-p_1-k)^{\beta}
(p_0-p_1-k)^{\nu}}{m_{\bar{D}^*}^2}\right] u_{\nu}(p_0, s_{P_c}) \Biggr\},
\end{eqnarray}
\begin{eqnarray}
\mM_{\bar{D} \Sigma_c^*-\pi} &=& i g_{\bar{D}^* \Sigma_c P_c}^{} g_{D^* D \pi}^{} g_{\pi \Sigma_c \Sigma_c^*}^{} \int_{-\infty}^{\infty} \frac{{\rm d}^4 k}{(2
\pi)^4} \Biggl\{ \tilde\Phi\left[ \frac{m_{\bar{D}^*}}{m_{\bar{D}^*} + m_{\Sigma_c}} p_0 - (p_1 + k) \right] \frac{\Lambda_1^4}{(m_{\pi}^2 - k^2)^2 + \Lambda_1^4} \notag \\
&& \mbox{} \qquad\qquad \bar{u}_{\mu}(p_1, s_{\Sigma_c^*}) k^\mu \frac{\slashed{p_1} + \slashed{k} + m_{\Sigma_c}}{(p_1+k)^2-m_{\Sigma_c}^2} \frac1{(p_0-p_1-k)^2-m_{\bar{D}^*}^2} \notag \\
&& \mbox{} \qquad \left[g^{\beta \nu} - \frac{(p_0-p_1-k)^{\beta} (p_0-p_1-k)^{\nu}}{m_{\bar{D}^*}^2}\right] \frac{-(p_0-p1)_\beta-k_\beta}{k^2-m_{\pi}^2} u_{\nu}(p_0, s_{P_c}) \Biggr\}, \\
\mM_{\bar{D} \Sigma_c^*-\rho} &=& -i g_{\bar{D}^* \Sigma_c P_c}^{} g_{D^* D \rho}^{} g_{\rho \Sigma_c \Sigma_c^*}^{} \int_{-\infty}^{\infty} \frac{{\rm d}^4 k}{(2
\pi)^4} \Biggl\{ \tilde\Phi\left[ \frac{m_{\bar{D}^*}}{m_{\bar{D}^*} + m_{\Sigma_c}} p_0 - (p_1 + k) \right] \frac{\Lambda_1^4}{(m_{\rho}^2 - k^2)^2 + \Lambda_1^4} \notag \\
&& \mbox{} \qquad \bar{u}^{\mu}(p_1, s_{\Sigma_c^*}) \gamma^\alpha \gamma5 \frac1{k^2-m_{\rho}^2}\left[k_\mu \left(g_{\alpha \beta} - \frac{k_{\alpha} k_{\beta}}{m_{\rho}^2}\right) - k_\alpha \left(g_{\mu \beta} - \frac{k_{\mu} k_{\beta}}{m_{\rho}^2}\right) \right] \notag \\
&& \mbox{} \qquad \frac{\slashed{p_1} + \slashed{k} + m_{\Sigma_c}}{(p_1+k)^2-m_{\Sigma_c}^2} \varepsilon^{\rho \sigma \lambda \beta} \left[(p_0-p_1-k)_\rho k_\lambda \right] \frac1{(p_0-p_1-k)^2-m_{\bar{D}^*}^2} \notag \\
&& \mbox{} \qquad \left[g_{\sigma \nu} - \frac{(p_0-p_1-k)_{\sigma} (p_0-p_1-k)_{\nu}}{m_{\bar{D}^*}^2}\right] \frac{-(p_0-p_1)_\beta-k_\beta}{k^2-m_{\pi}^2} u^{\nu}(p_0, s_{P_c}) \Biggr\},
\end{eqnarray}

\end{appendix}

% \newpage
\bigskip

\bibliographystyle{plain}

\end{document}